\documentclass[review]{elsarticle}
\usepackage{lineno,hyperref}
\usepackage{pgfplots}
\pgfplotsset{compat=newest}
\usepackage[utf8]{inputenc}
\usepackage{xcolor}
\usepackage{hyperref}
\usepackage{pgfplots}
\pgfplotsset{compat=1.15}
\usepackage{mathrsfs}
\usetikzlibrary{arrows}
\usepackage[flushleft]{threeparttable}
\usepackage{bigints}
\usepackage{verbatim}
\usepackage[edges]{forest}
\usepackage{placeins}

\usepackage{bbding}
\usepackage{cancel}
\usepackage{makecell}

\usetikzlibrary{arrows.meta, shapes.geometric, calc, shadows,fit}
\newcommand{\vect}[1]{\boldsymbol{#1}}

\colorlet{myPurple}{blue!40!red}
\colorlet{myGray}{gray!40!white}
\colorlet{Grayin}{white!20}
\colorlet{col6out}{gray!20!white}
\colorlet{col7in}{gray!20!white}
\colorlet{linecol}{black!90!white}

\usepackage{comment}
\usepackage{graphicx}
\usepackage{pgfplots}
\usetikzlibrary{spy}
\usepackage{amsmath}
\usepackage{multirow}
\usepackage{subcaption}
\usepackage{tabularx,array}
\newcolumntype{P}[1]{>{\raggedright\arraybackslash}p{#1}}
\newcolumntype{L}{>{\raggedright\arraybackslash}X}

\usepackage{framed} 
\usepackage{nomencl} 
\makenomenclature

\usepackage{etoolbox}
\renewcommand\nomgroup[1]{%
  \item[\bfseries
  \ifstrequal{#1}{A}{Acronyms}{%
  \ifstrequal{#1}{E}{English symbols}{%
    \ifstrequal{#1}{G}{Greek symbols}{%
        \ifstrequal{#1}{S}{Subscripts}{%
  \ifstrequal{#1}{U}{Superscripts}{}}}}}%
]}

\usepackage{upquote} 

\usepackage{xcolor}

\usepackage{enumitem}
\usepackage{tikz}
\usetikzlibrary{arrows.meta,
                chains,
                positioning,
                quotes,
                shapes.geometric}

\tikzset{FlowChart/.style={
startstop/.style = {rectangle, rounded corners, draw,
                    minimum width=3cm, minimum height=1cm, align=center,
                    on chain, join=by arrow},
  process/.style = {rectangle, draw,
                    text width=5cm, minimum height=1cm, align=center,
                    on chain, join=by arrow},
 decision/.style = {diamond, aspect=4, draw,
                    minimum width=3cm, minimum height=1cm, align=center,
                    on chain, join=by arrow},
                           io/.style = {trapezium, trapezium stretches body,   
                    trapezium left angle=70, trapezium right angle=110,
                    draw,
                    minimum width=3cm, minimum height=1cm,
                    text width =\pgfkeysvalueof{/pgf/minimum width}-2*\pgfkeysvalueof{/pgf/inner xsep},
                    align=center,
                    on chain, join=by arrow},
    arrow/.style = {thick,-Triangle}
                        }
        }

\usepackage{listings}

\usepackage{accsupp}
\newcommand{\noncopynumber}[1]{%
    \BeginAccSupp{method=escape,ActualText={}}%
    #1%
    \EndAccSupp{}%
}

\colorlet{mygray}{black!4}
\definecolor{mygreen}{rgb}{0,0.4,0}
\colorlet{myred}{red!80!blue}

\lstdefinestyle{cpp}{
	language=C++,
	basicstyle=\ttfamily\footnotesize,
	keepspaces=true,
  	columns=fixed,
	fontadjust=true,
	basewidth=0.5em,
	backgroundcolor=\color{mygray},
	tabsize=4,
	captionpos=b,
	frame=single,
	numbers=left,
	numberstyle=\tiny\noncopynumber,
	numbersep=5pt,
	breaklines=true,
	showstringspaces=false,
	keywordstyle=\color{blue},
	commentstyle=\color{mygreen},
	stringstyle=\color{myred}
}

\lstdefinestyle{OpenFOAMDict}{
	language=C++,
	basicstyle=\ttfamily\footnotesize,
	keepspaces=true,
  	columns=fixed,
	fontadjust=true,
	basewidth=0.5em,
	backgroundcolor=\color{mygray},
	tabsize=4,
	captionpos=b,
	frame=single,
	numbers=left,
	numberstyle=\tiny\noncopynumber,
	numbersep=5pt,
	breaklines=true,
	showstringspaces=false,
	commentstyle=\color{mygreen},
}

\usepackage[nameinlink,noabbrev]{cleveref}

\makeatletter
\def\@seccntformat#1{\@ifundefined{#1@cntformat}%
   {\csname the#1\endcsname\quad}
   {\csname #1@cntformat\endcsname}
}
\makeatother

\modulolinenumbers[50]

\journal{International Journal of Thermofluids}

\begin{document}
 \UseRawInputEncoding
\begin{frontmatter}


\title{Influence of Nanoparticle Shapes in Nanofluid Film Boiling on Vertical Cylinders: A Numerical Study}

\author{A. Yahyaee}
\ead{ayn@energy.aau.dk}

\address{AAU Energy, 9220-Aalborg East, Denmark}

\begin{abstract}
This investigation delves into the role of nanoparticle geometry in influencing the thermal and fluid dynamics characteristics of nanofluid film boiling along a vertical cylinder. Film boiling is a crucial process in specific industrial contexts, necessitating advanced thermal control and safety measures. Nanofluids are increasingly recognized for their potential to enhance these aspects. While the effects of nanoparticle type and concentration have been widely studied, this research provides a comprehensive examination of the impact of nanoparticle shape. The focus is on diverse shapes such as spheres, bricks, blades, cylinders, and platelets within alumina (Al$_2$O$_3$) water-based nanofluids.

Employing the Continuous-Species-Transfer (CST) approach within a Computational Multi-Fluid Dynamics (CMFD) framework, the study effectively models the intricate behaviors of nanoparticles with varied geometries in both liquid and vapor states. The incorporation of Brownian motion and thermophoretic dynamics enriches the analysis, enhancing the model’s fidelity.

A detailed 2D axisymmetric comparative analysis around a cylindrical surface is conducted. This allows for a thorough assessment of how different Al$_2$O$_3$ nanoparticle shapes affect parameters such as nanoparticle deposition, temperature gradients, and overall thermophysical properties. The study further explores the Nusselt number, a key indicator of heat transfer efficiency, providing valuable insights into the heat transfer enhancements attributable to varied nanoparticle shapes.

\end{abstract}



\begin{keyword}
Film Boiling\sep Nanofluids\sep Nanoparticle Shape\sep Computational Fluid Dynamics (CFD)\sep Computational Multi-Fluid Dynamics (CMFD)\sep Continuous-Species-Transfer (CST) Method\sep Brownian Motion\sep Thermophoresis\sep Vapor-Liquid Interface
\end{keyword}

\end{frontmatter}

\nomenclature[A]{VOF}{Volume of Fluid method, a technique for tracking fluid interfaces.}
\nomenclature[A]{CSF}{Continuum Surface Force model, used for simulating surface tension effects in fluids.}
\nomenclature[A]{CST}{Continuous-Species-Transfer model, employed for modeling species exchange between phases.}
\nomenclature[A]{CMFD}{Computational Multi-Fluid Dynamics, techniques for simulating multiple fluid phases.}

\nomenclature[E]{R}{Universal gas constant, applicable in ideal gas calculations.}
\nomenclature[E]{t}{Time variable in temporal analyses.}
\nomenclature[E]{$\mathbf{U}$}{Velocity vector of fluid flow.}
\nomenclature[E]{p}{Pressure within the fluid.}
\nomenclature[E]{$\mathbf{g}$}{Vector representing gravitational acceleration.}
\nomenclature[E]{$\dot{m}$}{Mass transfer rate between phases.}
\nomenclature[E]{T}{Temperature, indicating thermal state.}
\nomenclature[E]{c}{Specific heat at constant pressure.}
\nomenclature[G]{$\lambda$}{Thermal conductivity coefficient.}
\nomenclature[E]{h}{Enthalpy associated with phase change.}
\nomenclature[E]{M}{Molar mass of a substance.}
\nomenclature[E]{C}{Isothermal compressibility coefficient.}
\nomenclature[E]{Nu}{Nusselt number, representing convective heat transfer efficiency.}

\nomenclature[G]{$\rho$}{Density, indicative of mass per unit volume.}
\nomenclature[G]{$\vect{\tau}$}{Viscous stress tensor in fluids.}
\nomenclature[G]{$\alpha$}{Volume fraction of a phase in a mixture.}
\nomenclature[G]{$\gamma$}{Evaporation rate coefficient.}
\nomenclature[G]{$\mu$}{Dynamic viscosity, quantifying fluid resistance to deformation.}
\nomenclature[G]{$\kappa$}{Curvature measure at an interface.}
\nomenclature[G]{$\delta$}{Characteristic thickness of a layer or boundary.}

\nomenclature[S]{$\mathrm{p}$}{Indicative of a constant pressure condition.}
\nomenclature[S]{$\mathrm{l}$}{Referring to the liquid phase in a system.}
\nomenclature[S]{$\mathrm{saturat}$}{Indicates saturation conditions in thermodynamics.}

\nomenclature[E]{\(s\)}{Species concentration in vapor or liquid.}
\nomenclature[E]{\(S\)}{Total concentration in the entire domain.}
\nomenclature[E]{\(\phi\)}{Nanoparticle concentration in vapor or liquid.}
\nomenclature[E]{\(\varphi\)}{Total nanoparticle concentration in the system.}
\nomenclature[E]{$H_\mathrm{cnst}$}{Henry's law constant, defining solubility.}
\nomenclature[S]{$\infty$}{Condition at an infinitely far distance.}
\nomenclature[E]{d}{Characteristic dimension, typically diameter.}
\nomenclature[E]{$L_{\mathrm{fv}}$}{Latent heat of vaporization of a fluid.}
\nomenclature[S]{$\mathrm{np}$}{Refers to nanoparticles.}
\nomenclature[S]{$\mathrm{bf}$}{Indicates the base fluid in a mixture.}
\nomenclature[E]{k}{A constant, often in proportionality relations.}
\nomenclature[S]{$\mathrm{Bltzmn}$}{Related to the Boltzmann constant in thermodynamics.}
\nomenclature[E]{D}{Coefficient of diffusivity or diffusion.}
\nomenclature[S]{$\mathrm{Bn}$}{Associated with Brownian motion phenomena.}
\nomenclature[S]{$\mathrm{Thps}$}{Linked to Thermophoresis processes.}
\nomenclature[S]{$\mathrm{r}$}{Relates to a reference or baseline state.}
\nomenclature[G]{$\sigma$}{Surface tension coefficient in fluid mechanics.}
\nomenclature[E]{$\mathbf{f}$}{Force vector in a physical system.}
\nomenclature[S]{$\mathrm{grav}$}{Pertains to gravitational effects.}
\nomenclature[S]{$\mathrm{v,nf}$}{Refers to the vapor phase containing nanoparticles.}
\nomenclature[S]{$\mathrm{l,nf}$}{Denotes the liquid phase with nanoparticles.}
\nomenclature[S]{$\mathrm{c}$}{Related to compressive properties.}
\nomenclature[S]{$\alpha$}{Volume fraction of a specific phase.}
\nomenclature[S]{$\sigma$}{Refers to surface phenomena or interfacial properties.}
\nomenclature[S]{$\mathrm{f}$}{Referring to a facet or face in computational domains.}
\nomenclature[S]{$\mathrm{k}$}{Index for computational nodes.}
\nomenclature[S]{$\mathrm{num}$}{Denotes numerical results.}
\nomenclature[S]{$\mathrm{ana}$}{Indicates analytical results.}
\nomenclature[S]{$\delta$}{Represents thin films or layers.}
\nomenclature[S]{T}{Temperature in a thermal context.}
\nomenclature[S]{$\mathrm{Nu}$}{Nusselt number, important in heat transfer studies.}

\nomenclature[U]{*}{Signifies a dimensionless quantity.}
\nomenclature[U]{''}{Indicates a surface transfer rate.}
\nomenclature[U]{'''}{Denotes a volumetric transfer rate.}

\printnomenclature

\linenumbers

\section{Introduction}\label{Introduction}
\textbf{The Growing Importance of Nanofluid Film Boiling Research:}\\
Recent academic focus on enhancing thermal efficiency in nanofluid boiling processes underscores its diverse applications in sectors ranging from large vehicle propulsion and power generation to cooling systems in advanced electronics and aerospace thermal management. Prior studies have largely concentrated on pool \citep{Wen2023ExperimentalSurface,Dong2023TheSimulation} and flow boiling \citep{Joshi2023FlowNanorefrigerant,ElJery2023ComparisonNetwork} of nanofluids. Yet, the specific area of film boiling within nanofluids emerges as a critical, albeit less explored domain. Given the extreme thermal conditions in certain industrial applications, a thorough comprehension of film boiling is essential \citep{Wang2023MolecularSurfaces,Chen2023ASupply}. The incorporation of nanoparticles into fluids introduces additional complexity, warranting both computational \cite{Avramenko2021UnsteadyBoiling} and experimental investigations \citep{Santos2022EffectConditions}.

Following Maxwell's theoretical framework \citep{Maxwell1873AMagnetism}, adding particles to base fluids like water or oil can enhance thermal transfer properties \citep{Liang2019ReviewMicro-channels,Awais2021HeatReview}. However, microparticles face challenges such as sedimentation and clogging in conduits \citep{Qi2017ExperimentalTube}. Nanofluids, formulated by combining base liquids with nanoparticles, were introduced to overcome these limitations. These fluids exhibit unique thermal characteristics, offering advantages like improved thermal conductivity and changes in surface tension and dynamic viscosity \citep{Coccia2021ThermalEquation,Awais2021HeatReview,Riehl2022PerformancePipes}. Nonetheless, nanoparticle addition can sometimes reduce the effective heat capacity, necessitating careful evaluation for specific thermal applications.

\textbf{Impact of Nanoparticles on Surface Interactions and Fluid Dynamics in Nanofluid Boiling}

Nanoparticle characteristics substantially influence surface interactions and fluid dynamics in nanofluid boiling processes. One notable effect is on heated surfaces, where nanoparticle sedimentation occurs, affecting key factors like surface wettability, heat transfer coefficients, and capillary effects \citep{Borzuei2020RoleMotion,Wang2010OperationNanofluids,Esfahani2022TheStudy}. This sedimentation creates a layer that increases the effective contact area between phases, leading to improved heat transfer rates during phase change \citep{Rashidi2018ApplicationsReview}. Moreover, nanoparticles are instrumental in defining wettability, which consequently alters boiling behaviors. Critical parameters such as the triple line radius and contact angle are determined by the balance of forces among the liquid, gas, and solid at the three-phase interface \citep{Carey2020Liquid-vaporEquipment, Yin2019EffectsStudy}. The composition and concentration of nanoparticles are key in modifying surface tension and contact angles, with silica nanofluids, for example, exhibiting lower surface tension than water, thereby influencing boiling characteristics like bubble departure size \citep{Kujawska2021TheNanofluids, Vafaei2009TheNanofluids, Ally2010DetachmentBubbles}.

The impact of nanoparticles on droplet evaporation and bubble formation is also dependent on their size and type. Smaller nanoparticles tend to distribute more uniformly along the edge, while larger ones concentrate at the edges, affecting the evaporation dynamics \citep{Chon2007EffectDroplets}. Gold nanoparticles, in particular, have been observed to pin the triple line, significantly influencing bubble dynamics \citep{Vafaei2010TheLine}. This contrasts with other nanofluids where the triple line could expand significantly, up to 50 times the diameter of the nanoparticles, due to increased structural disjoining pressures, affecting both spreading behavior and bubble formation \citep{Chengara2004SpreadingGradient, Wasan2003SpreadingSolids,Nikolov2010NanoparticleSurface}.

\textbf{Comprehensive CFD Research in Nanofluid Boiling: From Basic Studies to Nanoparticle Movement Integration}

Advancements in Computational Fluid Dynamics (CFD) research related to nanofluid boiling have significantly progressed, with initial studies focusing on heat transfer enhancements and recent investigations integrating nanoparticle movement equations. Early research in this domain includes Abedini et al.'s exploration of subcooled nanofluid flow boiling \citep{Abedini2013NumericalNanofluid}, Shoghl et al.'s analysis of bubble dynamics in nanofluid pool boiling \citep{Shoghl2014ExperimentalBoiling}, and Qi et al.'s study on the effects of TiO2 nanoparticles in water-based nanofluids \citep{Qi2017ExperimentalTube}. Abdollahi et al. examined the influence of various nanoparticle types and concentrations on laminar boiling in micro-channels \citep{Abdollahi2017FluidArrangement}, while Mohammed et al. confirmed the significant effect of nanoparticle concentration on heat transfer in diverse boiling scenarios \citep{Mohammed2018CFDTube,Mohammed2020ThermalStudy}.

Further developments include Soleimani et al.'s observations of improved thermal management with HFE-7100 nanofluids \citep{Soleimani2020ThermalNanofluid}, and Li et al.'s investigation into the gravitational effects on boiling fluids \citep{Li2021NumericalDirections}. Pournaderi et al. focused on Al2O3-water nanofluid droplets, noting enhanced heat extraction at specific concentrations \citep{Pournaderi2021ModelingSurface}. Zaboli et al. and Alomar et al. extended research to nanoparticle concentrations in silica and salt solution nanofluids \citep{Zaboli2022NumericalConcentrations,Alomar2022ModellingNano-fluid}, while Bahiraei et al. provided insights into entropy generation and exergy in Ag-water nanofluid flow \citep{Bahiraei2022Two-phaseConsiderations}.

Moving towards a more holistic approach, recent CFD research initiated by Buongiorno \citep{Buongiorno2006ConvectiveNanofluids} has begun incorporating Brownian diffusion and thermophoresis as primary transport mechanisms for nanoparticles. This direction is further explored by Yang et al. \citep{Yang2013ConvectiveAnnulus}, Malvandi et al. \citep{Malvandi2014ModifiedPipe,Malvandi2016ThermophoresisCylinder}, and Hedayati et al. \citep{Hedayati2015EffectsMicrochannel,Hedayati2016NanoparticleMicrochannel} across various nanofluids and boiling scenarios. Michaelides \citep{Michaelides2015BrownianLiquids} and Lin et al. \citep{Lin2018EffectsSimulation} expanded on these findings, while Malvandi \citep{Malvandi2016AnisotropicField} delved into the behavior of magnetic nanofluids. Wang et al.'s numerical study \citep{Wang2023NumericalSystem} on nanoparticle deposition in microchannel cooling systems further highlights the importance of nanoparticle movement in nanofluid thermal management. These cumulative efforts underscore the evolving understanding of nanofluid boiling dynamics, bridging foundational studies with advanced nanoparticle movement considerations.

\textbf{Experimental and Numerical Investigations on the Influence of Nanoparticle Shape in Nanofluid Boiling}

The current research landscape reveals a notable gap in understanding the specific influence of nanoparticle shapes on the thermal and hydraulic performance during nanofluid boiling. This section aims to address this gap by reviewing both experimental and numerical studies that scrutinize the impact of nanoparticle shapes on single-phase or two-phase cooling systems.

Previous studies have underscored the significant role of nanoparticle characteristics, such as shape, size, and volume fraction, on the heat transfer properties of nanofluids \citep{Hwang2009FlowNanoparticles,Yu2011LaminarNanoparticles,Ferrouillat2013InfluenceNanofluids,Timofeeva2009ParticleNanofluids,Merilainen2013InfluenceNanofluids}. Notably, water-based Al$_2$O$_3$ nanofluids demonstrated a substantial increase in convective heat transfer coefficient, with variations attributed to nanoparticle shape \citep{Hwang2009FlowNanoparticles}. The impact of nanoparticle properties, like aspect ratio and aggregation, is particularly pronounced in nanofluids with non-spherical particles \citep{Yu2011LaminarNanoparticles}. Studies by Ferrouillat et al. and Timofeeva et al. indicated that different nanoparticle shapes and sizes could alter the Nusselt number and heat transfer coefficients \citep{Ferrouillat2013InfluenceNanofluids,Timofeeva2009ParticleNanofluids}. Moreover, an increase in nanoparticle volume fraction beyond 2\% presented a trade-off between enhanced heat transfer and decreased efficiency due to pressure losses, with factors like nanoparticle dispersion and thermophoresis playing a role \citep{Merilainen2013InfluenceNanofluids,Chandrasekar2012MechanismsReview}.

While the majority of theoretical studies have employed conventional single-phase models, these often deviate from experimental data due to overlooking the slip mechanisms between nanoparticles and the base fluid \citep{Vanaki2016NumericalReview}. Elias et al. demonstrated that cylindrical nanoparticles were more effective in enhancing heat transfer in shell and tube heat exchangers compared to other shapes \citep{Elias2013EffectExchanger}. Conversely, Vanaki et al. found that platelet-shaped nanoparticles led to the highest heat transfer enhancement \citep{Vanaki2014EffectShifts}.

Two-phase models, considering both solid and liquid phases, provide a deeper insight into nanofluid behavior. Bahremand et al.'s use of a Lagrangian--Eulerian two-phase approach for modeling turbulent convection flow of water--silver nanofluid with spherical nanoparticles yielded more accurate predictions than homogeneous models \citep{Bahremand2015ExperimentalApproach}. However, there is a dearth of literature employing two-phase models to study the effect of nanoparticle shapes on convective heat transfer. Liu et al. introduced new shape descriptors, flatness, and elongation, in their two-phase model studies, finding that platelet-shaped nanoparticles enhanced heat transfer in curved square ducts the most \citep{Liu2018EffectsModel}. Similarly, Jiang et al.'s recent work utilizing a two-phase mixture model revealed that blade-shaped nanoparticles outperformed spherical ones in minichannels, enhancing the average Nusselt number by 8.3\% at a nanoparticle volume fraction of 0.05 \citep{Jiang2020EffectMinichannel}.

\section*{Inspirations and Contributions}

The review of existing literature reveals several pivotal gaps, which form the foundation for the current study:

\begin{itemize}
\item Despite Buongiorno's detailed analysis of forces such as Brownian diffusion and thermophoresis affecting nanoparticles \citep{Buongiorno2006ConvectiveNanofluids}, a considerable part of CFD research \citep{Abedini2013NumericalNanofluid, Abdollahi2017FluidArrangement, Shoghl2014ExperimentalBoiling, Qi2017ExperimentalTube, Mohammed2018CFDTube, Mohammed2020ThermalStudy, Soleimani2020ThermalNanofluid, Li2021NumericalDirections, Zaboli2022NumericalConcentrations, Alomar2022ModellingNano-fluid, Mavi2022Volume-of-FluidChannels} still lacks the incorporation of dedicated nanoparticle transport equations. Moreover, even in CFD studies that do include these equations \citep{Yang2013ConvectiveAnnulus, Malvandi2014ModifiedPipe, Malvandi2014EffectsChannel, Hedayati2015EffectsMicrochannel, Hedayati2016NanoparticleMicrochannel, Michaelides2015BrownianLiquids, Malvandi2016ThermophoresisCylinder, Lin2018EffectsSimulation}, there is a tendency to focus predominantly on the insulating role of the vapor layer in film boiling, often overlooking the critical interaction between the nanofluid’s liquid phase and the vapor layer.

\item Despite extensive CFD and experimental research, the effect of nanoparticle shapes on thermal and hydraulic performance during boiling is yet to be fully explored.
\end{itemize}

In response to these gaps, this study introduces a CFD analysis of film boiling in nanofluids, emphasizing the effect of various nanoparticle shapes like spheres, bricks, blades, cylinders, and platelets in water-based nanofluids. The research pays special attention to the diffusion behaviors of these diverse nanoparticle shapes between the vapor and liquid phases, as well as their concentration variations at the phase interface.

The CFD approach utilizes the Continuous-Species-Transfer (CST) methodology, an advancement in Computational Multi-Fluid Dynamics (CMFD). This model provides an accurate simulation of crucial aspects in film boiling, particularly focusing on nanoparticle deposition on the heated surface and the transfer of nanoparticles between the vapor and liquid phases. Film boiling is the chosen study regime as it involves direct contact between the vapor layer and the heated solid surface, with the liquid phase remaining separated. This aspect minimizes the effects of changes in the triple line during nanofluid boiling. Analytical considerations involving Brownian motion and thermophoresis are incorporated to validate the findings, underscoring the CST methodology's effectiveness in simulating nanofluid film boiling.

The primary aim is to perform a detailed CFD analysis on film boiling of nanofluids, with a unique emphasis on the role of nanoparticle shapes. This computational model is designed to elucidate complex behaviors at the vapor-liquid interface during film boiling in nanofluids, propelling both the subject matter and methodology into new territories of investigation.

\section{Mathematical Modeling Approach}

\subsection{Formulation of Governing Equations}

The current research employs the Volume of Fluid (VOF) method to analyze the dynamics of a two-phase system, specifically focusing on the phase transitions involving thermal changes. This method bases its analysis on a unified set of conditionally volume-averaged conservation laws, applicable to both the liquid and vapor phases of the nanofluid. The core equations that govern the system dynamics are presented as follows:

The principle of mass conservation is encapsulated in the equation:
\begin{equation}
\cfrac{\partial \rho}{\partial t} + \nabla \cdot (\rho \mathbf{U}) = 0,
\label{eq:mass_conservation_refined}
\end{equation}

For momentum conservation, the equation is represented as:
\begin{equation}
\cfrac{\partial (\rho \mathbf{U})}{\partial t} + \nabla \cdot (\rho \mathbf{U U}) = -\nabla p + \nabla \cdot \boldsymbol{\tau} + \mathbf{f}_{\text{grav}} + \mathbf{f}_{\sigma}.
\label{eq:momentum_conservation_refined}
\end{equation}

In the model, the density $\rho$ and other variables related to the mixture are considered as functions of volumetric mass concentrations, accounting for both the nanofluid and vapor phases. Surface tension forces are addressed using the Continuous-Surface-Force (CSF) model \citep{Brackbill1992ATension}, formulated as:
\begin{equation}
\mathbf{f}_{\sigma} = \sigma \kappa \nabla \alpha_{\text{l,nf}}, \quad \text{where} \quad \kappa = -\nabla \cdot \left( \cfrac{\nabla \alpha_{\text{l,nf}}}{|\nabla \alpha_{\text{l,nf}}|} \right),
\end{equation}

In this context, $\alpha_{\text{l,nf}}$ denotes the liquid nanofluid's volume fraction in a computational cell. The phase fraction evolution is given by:
\begin{equation}
\cfrac{\partial \alpha_{\text{l,nf}}}{\partial t} + \nabla \cdot (\mathbf{U} \alpha_{\text{l,nf}}) + \nabla \cdot (\mathbf{U}_\mathrm{r} \alpha_{\text{l,nf}} (1 - \alpha_{\text{l,nf}})) = -\dot{m}'''\left[ \cfrac{1}{\rho_{\text{l,nf}}}-\alpha_{\text{l,nf}} \left (\cfrac{1}{\rho_{\text{l,nf}}}-\cfrac{1}{\rho_{\text{v,nf}}}\right )\right],
\label{eq:phase_fraction_change}
\end{equation}

The relative velocity \( \mathbf{U}_r \)  at a distinct interface is typically modeled as compressive velocity  \( \mathbf{U}_c \), perpendicular to the interface, following the guidelines by Weller and Olsson \citep{Weller2006AFlow,Olsson2005AFlow}:
\begin{equation}
\mathbf{U}_c = \min \left[ c_{\alpha} | \mathbf{U} |, \max (| \mathbf{U} |) \right] \cfrac{\nabla \alpha_{\text{l,nf}}}{| \nabla \alpha_{\text{l,nf}} |},
\label{eq:compressive_velocity}
\end{equation}

This modeling technique is designed to maintain a distinct delineation of the phase interface, with \( c_\alpha \) set between 1 and 4, to minimize numerical diffusion errors at the interface.

\textbf{Adaptation of Governing Equation for Nanoparticle Concentrations}\\
Regarding the equation for nanoparticle concentrations, the foundation of this study is built upon a key governing equation, referenced in several seminal studies \citep{Buongiorno2006ConvectiveNanofluids,Yang2013ConvectiveAnnulus,Malvandi2014EffectsChannel,Avramenko2021UnsteadyBoiling,Lin2018EffectsSimulation}. Traditionally applied in single-phase (either vapor or liquid) analysis, this equation forms the bedrock of the computational approach used:

\begin{equation}
\cfrac{\partial \phi}{\partial t} + \nabla \cdot (\phi \mathbf{U}) = \nabla \cdot (D_{\text{Bn}} \nabla \phi) + \nabla \cdot\left[D_{\text{Thps}} \cfrac{\nabla T}{T_{\infty}}\right].
\label{eq:nano_concentration_basic}
\end{equation}

To accurately model nanoparticle behavior during film boiling in nanofluids, this research extends the analysis to encompass both vapor and liquid phases and their interfacial dynamics. By integrating the Continuous-Species-Transfer (CST) method \citep{Marschall2012NumericalOpenFOAM}—with its formulation detailed in Appendix \ref{apendixNanoCon}—an enhanced governing equation is presented. This equation comprehensively encapsulates the diverse transport phenomena and complexities at the fluid interface:

\begin{align}
    \cfrac{\partial \varphi}{\partial t} + \nabla \cdot \left( \varphi \mathbf{U} \right) &= \nabla \cdot \left( D_\mathrm{Bn}'T \nabla \varphi \right) + \nabla \cdot \left( \varphi T \nabla D_\mathrm{Bn}'  \right)+ \nabla \cdot \left( \varphi D_\mathrm{Bn}' \nabla T  \right) \nonumber \\
    &\quad - \nabla \cdot \left[ \left( \cfrac{\overline{D_\mathrm{Bn}' }^{\mathrm{l,nf}} - \overline{D_\mathrm{Bn}' }^{\mathrm{v,nf}} / \mathrm{H}_\mathrm{cnst}}{\alpha_{\mathrm{l,nf}} + (1 - \alpha_{\mathrm{l,nf}}) / \mathrm{H}_\mathrm{cnst}} \right) \varphi T \nabla \alpha_{\mathrm{l,nf}} \right] \nonumber \\
    &\quad - \nabla \cdot \left[ \left( \cfrac{\alpha_{\mathrm{l,nf}}\overline{D_\mathrm{Bn}' }^{\mathrm{l,nf}} +(1 - \alpha_{\mathrm{l,nf}}) \overline{D_\mathrm{Bn}' }^{\mathrm{v,nf}} / \mathrm{H}_\mathrm{cnst}}{\alpha_{\mathrm{l,nf}} + (1 - \alpha_{\mathrm{l,nf}}) / \mathrm{H}_\mathrm{cnst}} \right) \varphi \nabla T \right]\nonumber \\
    &\quad + \nabla \cdot \left[ \left( \cfrac{\alpha_{\mathrm{l,nf}}\overline{D_\mathrm{Thps}' }^{\mathrm{l,nf}} +(1 - \alpha_{\mathrm{l,nf}}) \overline{D_\mathrm{Thps}' }^{\mathrm{v,nf}} / \mathrm{H}_\mathrm{cnst}}{\alpha_{\mathrm{l,nf}} + (1 - \alpha_{\mathrm{l,nf}}) / \mathrm{H}_\mathrm{cnst}} \right) \varphi \nabla T \right].
    \label{eq:final_governing_equationAA}
\end{align}

Equation \ref{eq:final_governing_equationAA} thoroughly addresses the entire computational domain. It not only represents the general transport of nanoparticles but also precisely captures the complex dynamics occurring at the fluid interface.

\textbf{Adaptation of Energy Equation}

This research builds upon the fundamental energy conservation equation, modified to integrate nanoparticle influences:

\begin{equation}
\cfrac{\partial (\rho c_{\text{p}} T)}{\partial t} + \nabla \cdot (\rho c_{\text{p}} \mathbf{U} T) = \nabla \cdot (\lambda \nabla T) - \dot{m}''' L_{\text{fv}},
\end{equation}

The inclusion of Brownian motion and thermophoresis effects, following guidance from Malvandi and others \cite{Malvandi2016ThermophoresisCylinder}, introduces additional terms:

\begin{equation}
+\rho_{\text{np}} c_{\text{p,np}}\left[D_{\text{Bn}} \nabla \phi \cdot \nabla T+\cfrac{D_{\text{Thps}}}{T_{\infty}} \nabla T \cdot \nabla T\right],
\label{eq:tempBrownianModified}
\end{equation}

The original formulation, given in Equation~\ref{eq:tempBrownianModified}, initially accounts for a single phase. To extend its applicability across the complete domain, including vapor, liquid, and their interface, modifications are made as detailed in Appendix~\ref{apendixNanoCon}. The revised thermal energy conservation equation, accounting for all phases and the influence of nanoparticle movement, is thus:

\begin{align}
\cfrac{\partial (\rho c_{\text{p}} T)}{\partial t} + \nabla \cdot (\rho c_{\text{p}} \mathbf{U} T) &= \nabla \cdot (\lambda \nabla T) - \dot{m}''' L_{\text{fv}} \nonumber \\
&+\rho_{\text{np}} c_{\text{p,np}} \left( \cfrac{\alpha_{\text{l,nf}}\overline{D_{\text{Bn}}' }^{\text{l,nf}} +(1 - \alpha_{\text{l,nf}}) \overline{D_{\text{Bn}}' }^{\text{v,nf}} / H_{\text{cnst}}}{\alpha_{\text{l,nf}} + (1 - \alpha_{\text{l,nf}}) / H_{\text{cnst}}} \right) T \nabla \varphi \nabla T  \nonumber \\
&+\rho_{\text{np}} c_{\text{p,np}}\left( \cfrac{\alpha_{\text{l,nf}}\overline{D_{\text{Thps}}' }^{\text{l,nf}} +(1 - \alpha_{\text{l,nf}}) \overline{D_{\text{Thps}}' }^{\text{v,nf}} / H_{\text{cnst}}}{\alpha_{\text{l,nf}} + (1 - \alpha_{\text{l,nf}}) / H_{\text{cnst}}} \right) \varphi \nabla T \nabla T
.
\label{eq:finalEnergyConservation}
\end{align}

Utilizing OpenFOAM, a comprehensive open-source CFD tool \cite{Weller2006AFlow}, the research models the phase change process to accurately determine the mass flux rate \( \dot{m}''' \). The interfaceHeatResistance model in OpenFOAM v2006, based on concepts by Hardt and others \cite{Hardt2008EvaporationTerms}, simplifies this representation:

\begin{equation}
\dot{m}'' = \cfrac{2 - \gamma}{2\gamma} \sqrt{\cfrac{M}{2\pi R}} \cfrac{\rho_{\text{v,nf}} L_{\text{fv}}(T - T_{\text{saturat}})}{(T_{\text{saturat}})^{3/2}},
\end{equation}

Here, \( \gamma \) represents coefficients related to evaporation, and \( R \) is the universal gas constant (approximately \( 8.314 \, \text{J/(mol K)} \)), normalizing the molecular weight \( M \). The volumetric mass transfer rate is then:

\begin{equation}
\dot{m}''' = \dot{m}'' |\nabla \alpha_{\text{l,nf}}|.
\end{equation}

\subsection{Description of the Problem}

\begin{figure}[!htb]
	\centering
\begin{tikzpicture}[line cap=round,line join=round,>=triangle 45]
\fill[fill=blue!20] (0,0) parabola (0.5,1) parabola bend (0.5,1) (1,12)--(4,12)--(4,0);
\draw[dashed](0,0) parabola (0.5,1) parabola bend (0.5,1) (1,12);
\draw (0,0) -- (4,0) -- (4,12) -- (0,12) -- (0,0);
\filldraw[fill=gray!20] (-6,0) -- (-6,12) arc (180:360:3cm and 0.25cm) -- (0,0) arc (360:180:3cm and 0.25cm);
\draw[dashed] (-3,0) ellipse (3 and 0.25);
\draw (-3,12) ellipse (3 and 0.25);
\draw[->] (2,6) -- (2,5) node[anchor=south west] {$\vect{g}$};
\node at (-2.3,5.4) {$\cfrac{\partial P}{\partial r} = \alpha_\mathrm{l,nf} = \mathbf{U} = 0$};
\node at (-2.6,3.2) {$\left(D_\mathrm{Bn} \cfrac{d\varphi}{dr} \right)  = - \left( \cfrac{D_\mathrm{Thps}}{T_\infty} \cfrac{dT}{dr} \right)$};
\node at (-2.3,4.4) {$T = T_{\text{saturat}} + 4$};
\node at (4.8,6) {$\cfrac{\partial T}{\partial r} =   0$};
\node at (5,5) {$\cfrac{\partial \alpha_\mathrm{l,nf}}{\partial r} =0$};
\node at (4.9,4) {$\cfrac{\partial \mathbf{U}}{\partial r} =0$};
\node at (4.95,3) {$P = P_{\text{saturat}}$};
\node at (2,-1) { $T = T_{\text{saturat}} \; \;\;    P = P_{\text{saturat}} \; \;\;\alpha_\mathrm{l,nf} = 1 \; \;\; \cfrac{\partial \mathbf{U}}{\partial x} = 0$ };
\node at (1.75,13.3) {$    \cfrac{\partial P}{\partial x} = \cfrac{\partial \mathbf{U}}{\partial x} = \cfrac{\partial T}{\partial x} = 0 \; \;\;\alpha_\mathrm{l,nf} = 1 $};
\end{tikzpicture}
	\caption{Schematic depicting the boundary constraints for laminar film boiling along a vertical interface. }
	\label{sec2fig:vertical}
\end{figure}
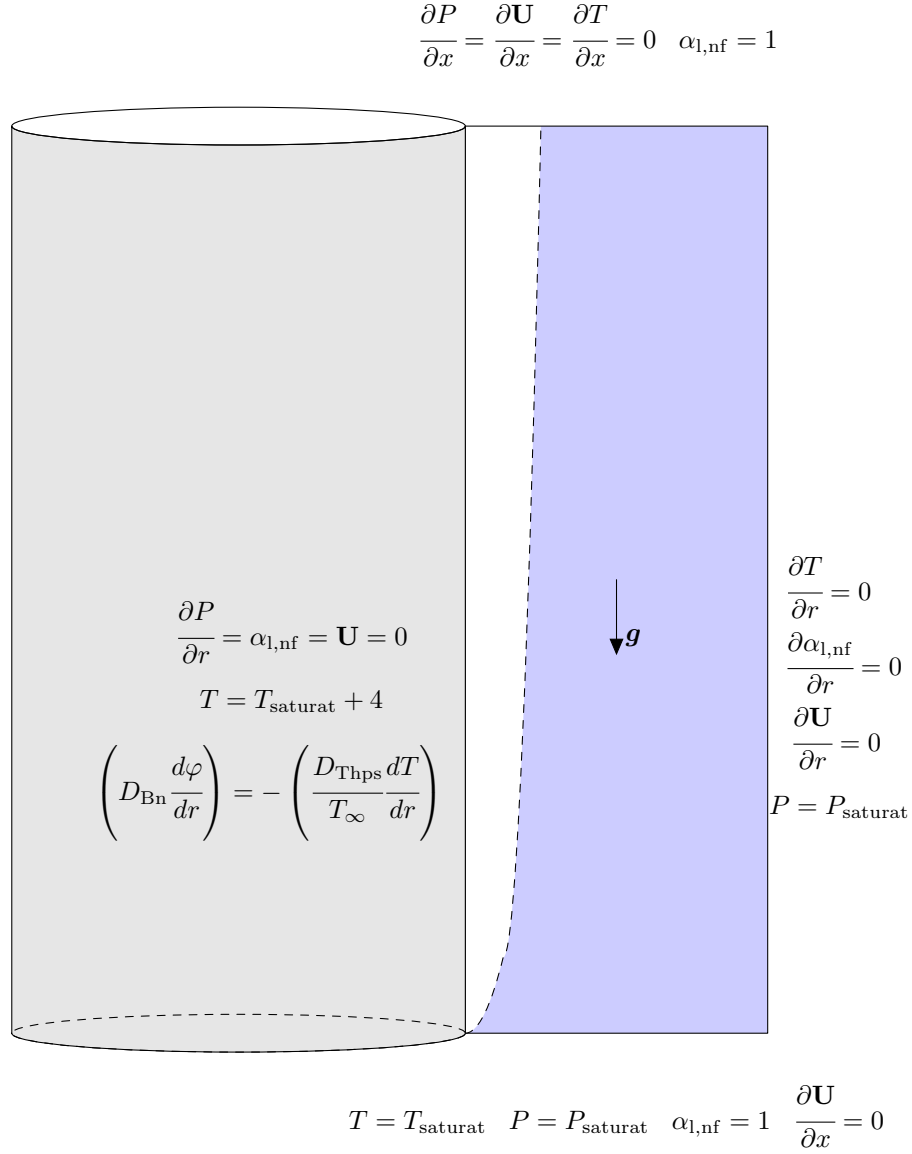

The problem under investigation is depicted in Figure~\ref{sec2fig:vertical}, where a vertical cylinder is immersed in a nanofluid. The process commences when the cylinder's surface temperature, \( T_{\text{s}} \), surpasses the boiling point, \( T_{\text{saturat}} \). This leads to the formation of a vapor layer enveloping the cylinder, extending vertically along its length. This vapor layer's ascent along the cylinder's z-axis is influenced by buoyancy. In the model, the movement of the vapor is primarily influenced by the force of gravity. 

Here, the complexity of the three-dimensional (3D) scenario is reduced to a two-dimensional (2D) axisymmetric representation for computational efficiency. The domain is configured as a wedge, where the z-axis, located at the cylinder's vertical midpoint, acts as the axis of rotation. In this setup, the front and back boundaries of the domain are treated as 'wedge' types in OpenFOAM. This designation is crucial for simulating axisymmetric physical phenomena with enhanced computational efficiency.

The numerical model initiates with a thin vapor film in direct contact with the surface of the cylinder. The prescribed boundary conditions for this simulation are depicted in Figure~\ref{sec2fig:vertical}. The cylinder's surface, representing the left boundary in the context of the simulation, is governed by a concentration boundary condition for the nanoparticles as encapsulated by a specific equation, seen in Figure~\ref{sec2fig:vertical}. The coefficients for thermophoretic (\(D_{\text{Thps}}\)) and Brownian diffusion (\(D_{\text{Bn}}\)) in that specific equation are quantified by Equations~\ref{eq:brownianD} and~\ref{eq:thermophoresisD}, respectively. A fundamental premise of the boundary condition at the cylinder wall is the impermeability to nanoparticles, which stipulates a null net flux of nanoparticles. This zero-flux condition contemplates the collective migration of nanoparticles, often referred to as Stefan flow, which is responsive to both concentration and temperature differentials. This concept is grounded in the investigations of Avramenko et al., as elaborated in their studies \citep{Avramenko2011Self-similarLayer, Avramenko2012SymmetryNanofluid, Malvandi2016ThermophoresisCylinder}.

\subsection{Nanoparticles Shapes and Thermophysical Properties}

\begin{figure}[ht]
\centering
\begin{tikzpicture}[baseline]

    \node at (0,0) [circle, shade, ball color = gray!40, opacity = 0.4, minimum size = 1cm] {};
    \node[below] at (0,-1) {Sphere}; 

    \node at (3,0) [ellipse, fill=teal!20, minimum width=2.5cm, minimum height=0.3cm] {};
    \node[below] at (3,-1) {Blade}; 

    \begin{scope}[shift={(6,0)}]
        \filldraw[fill=yellow!20!white, draw=black] (-0.5,-0.5) rectangle (0.5,0.5); 
        \draw (-0.45, -0.45) -- (-0.45, 0.55) -- (0.55, 0.55) -- (0.55, -0.45) -- cycle; 
        \draw (-0.5, 0.5) -- (-0.45, 0.55);
        \draw (0.5, 0.5) -- (0.55, 0.55);
        \draw (0.5, -0.5) -- (0.55, -0.45);
        \node[below] at (0,-1) {Brick}; 
    \end{scope}

    \node at (9,0) [cylinder, shape border rotate=90, draw, fill=purple!20, minimum height=1cm, minimum width=0.6cm] {};
    \node[below] at (9,-1) {Cylinder}; 

    \begin{scope}[shift={(12,0)}]
        \filldraw[fill=gray!20] (-0.75,0) -- (-0.75,0.25) arc (180:360:0.75cm and 0.25cm) -- (0.75,0) arc (360:180:0.75cm and 0.25cm); 
        \draw (0,0.25) ellipse (0.75 and 0.25);
        \node[below] at (0,-1) {Platelet}; 
    \end{scope}

\end{tikzpicture}
\caption{Illustration of different nanoparticle shapes}
\label{fig:nanoparticle_shapes}
\end{figure}
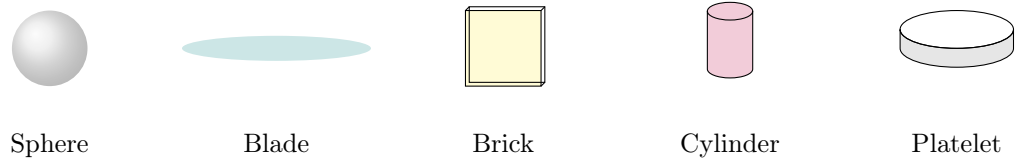

In the current analysis, a diverse array of nanoparticle geometries are incorporated, encompassing spheres, blades, bricks, cylinders, and platelets, as delineated in Figure~\ref{fig:nanoparticle_shapes}. Each nanoparticle shape is standardized to have an equivalent spherical diameter (\(d_p\)) of 45 nm for consistency. The term 'equivalent spherical diameter' refers to the diameter of a sphere that embodies the same volume as the irregularly shaped particle. The detailed geometrical specifications for each nanoparticle form are based on the seminal definitions provided by the source \cite{Jiang2020EffectMinichannel}.

For the numerical simulations to yield reliable results, it is crucial to employ accurate models for the nanofluid's thermophysical properties. Contrary to the simpler rule of mixture, which has been deemed inadequate by the findings of Das et al. \cite{Das2006HeatReview}, a more sophisticated approach is required to compute these properties for nanofluids. In the present investigation, the computational formulae for calculating the nanofluid's density and specific heat capacity are as follows:

\begin{equation}
    \rho_{\text{nf}} = \varphi \rho_{\text{np}} + (1 - \varphi) \rho_{\text{bf}},
    \label{eq:modified_density}
\end{equation}
\begin{equation}
    c_{\text{p,nf}} = \cfrac{\varphi \rho_{\text{np}} c_{\text{p,np}} + (1 - \varphi) \rho_{\text{bf}} c_{\text{p,bf}}}{\rho_{\text{nf}}},
    \label{eq:modified_heat_capacity}
\end{equation}
where \(\varphi\) represents the volume fraction of nanoparticles in the nanofluid. The notations 'np', 'bf', and 'nf' stand for nanoparticles, base fluid, and nanofluid, respectively.

For the evaluation of nanofluid thermal conductivity with various nanoparticle shapes, the model aligns with the Hamilton-Crosser framework, as outlined in Equation~\ref{eq:revised_thermal_conductivity} \cite{Hamilton1962ThermalSystems}:

\begin{equation}
\lambda_{\text{nf}} = \lambda_{\text{bf}} \cfrac{\lambda_{\text{p}} + (n - 1)\lambda_{\text{bf}} + (n - 1)(\lambda_{\text{p}} - \lambda_{\text{bf}})\varphi}{\lambda_{\text{p}} + (n - 1)\lambda_{\text{bf}} - (\lambda_{\text{p}} - \lambda_{\text{bf}})\varphi}
\label{eq:revised_thermal_conductivity}
\end{equation}

Here, $n$ represents the empirical shape factor, defined as $n = \cfrac{3}{\Psi}$, where $\Psi$ is the nanoparticle's sphericity, calculated as the ratio of the surface area of an equivalent volume sphere to the nanoparticle's actual surface area. The sphericity values for different nanoparticle shapes are presented in Table~\ref{table:sphericity_values}.

\begin{table}[ht]
    \centering
    \caption{Sphericity of different shaped nanoparticles ($d_{\text{p}} = 45$ nm)}
    \label{table:sphericity_values}
    \begin{tabular}{|c|c|c|c|c|c|}
        \hline
        & Sphere & Brick & Cylinder & Platelet & Blade \\
        \hline
        Sphericity ($\Psi$) & 1 & 0.81 & 0.62 & 0.52 & 0.36 \\
        \hline
    \end{tabular}
\end{table}

The computation of the nanofluid's dynamic viscosity for spherical nanoparticles utilizes Equation~\ref{eq:revised_dynamic_viscosity_spherical} \cite{Brinkman1952TheSolutions}:

\begin{equation}
\mu_{\text{nf}} = \cfrac{\mu_{\text{f}}}{(1 - \varphi)^{2.5}}
\label{eq:revised_dynamic_viscosity_spherical}
\end{equation}

For nanofluids containing non-spherical nanoparticles, the dynamic viscosity is determined by Equation~\ref{eq:revised_dynamic_viscosity_nonspherical} \cite{Timofeeva2009ParticleNanofluids}:

\begin{equation}
\mu_{\text{nf}} = 1 + A1 \cdot \varphi + A2 \cdot \varphi^2
\label{eq:revised_dynamic_viscosity_nonspherical}
\end{equation}

The coefficients $A1$ and $A2$ are specific to non-spherical nanoparticles and are detailed in Table~\ref{table:viscosity_coefficients}.

\begin{table}[ht]
    \centering
    \caption{Viscosity coefficients for different shaped nanoparticles}
    \label{table:viscosity_coefficients}
    \begin{tabular}{|c|c|c|c|c|}
        \hline
        & Platelet & Blade & Cylinder & Brick \\
        \hline
        A1 & 37.1 & 14.6 & 13.5 & 1.9 \\
        \hline
        A2 & 612.6 & 123.3 & 904.4 & 471.4 \\
        \hline
    \end{tabular}
\end{table}

The thermophysical properties of the base fluid and its vapor phase, as well as those of the Al\(_2\)O\(_3\) nanoparticles, which are used in this study, are elaborated in Tables~\ref{vertTable} and~\ref{aluminaTable}, laying the groundwork for the numerical model.

\begin{table}[!htb]
\centering
\caption{Properties of saturated water at 21.9 MPa and 346 K, used in modeling the 2D film boiling and 2D axisymmetric vertical film boiling scenarios.}
\label{vertTable}
\begin{threeparttable}
\begin{tabular}{cccc}
\hline
& \textbf{Dimension} & \textbf{Base Fluid} & \textbf{Vapor} \\
\hline
Thermal conductivity, $\lambda$ & $\mathrm{W}\, \mathrm{m}^{-1}\, \mathrm{K}^{-1}$ & 0.545 & 0.054 \\
Density, $\rho$ & $\mathrm{kg}\, \mathrm{m}^{-3}$ & 402.4 & 242.7 \\
Viscosity, $\mu$ & Pa s & $4.53 \times 10^{-5}$ & $3.23 \times 10^{-5}$ \\
Specific heat capacity $\mathrm{c}_{p}$ & $\mathrm{J}\,\mathrm{kg}^{-1}\,\mathrm{K}^{-1}$ & $2.18\times10^5$ & $3.52\times10^5$ \\
Latent heat, $h$ & $\mathrm{J}\,\mathrm{kg}^{-1}$ & $2.24\times10^6$ & $1.96\times10^6$ \\
Surface tension, $\sigma$ & $\mathrm{N}\,\mathrm{m}^{-1}$ & $7 \times 10^{-5}$ & - \\
\hline
\end{tabular}
\end{threeparttable}
\end{table}

\begin{table}[!htb]
\caption{Thermophysical properties of Al\(_2\)O\(_3\) nanoparticles. \label{aluminaTable}}
\setlength{\tabcolsep}{6mm}
\centering
\begin{tabular}{ccc}
    \hline
    \textbf{Property} & \textbf{Units} & \textbf{Al\(_2\)O\(_3\)} \\
    \hline
    Thermal conductivity, \(\lambda\) & W\,m\(^{-2}\)\,K\(^{-1}\) & 40 \\
    Density, \(\rho\) & kg\,m\(^{-3}\) & 3970 \\
    Specific heat capacity, \(c_p\) & J\,kg\(^{-1}\)\,K\(^{-1}\) & 765 \\
    \hline
\end{tabular}
\end{table}

\subsection{Benchmark Validation via Taylor's Two-Dimensional Film Boiling Model}

\begin{figure}[!htb]
	\centering
\begin{tikzpicture}[line cap=round,line join=round,>=triangle 45,x=1cm,y=.85cm]
\begin{scope}[scale=.45]
\clip(-5,-4) rectangle (20,32);
\draw[smooth,samples=100,domain=1.1:5.6] plot(\x,{sin((0.7*(\x)+3.9)*180/pi)});
\draw[line width=0.1pt]  (10,29)  (10,24.5)  (0,24.5)  (0,29)--(10,29);
\draw[line width=0.1pt]  (10,-1.5) -- (10,24.5) -- (0,24.5) -- (0,-1.5)--(10,-1.5);

\draw[smooth,samples=100,domain=1.1:5.6] plot(\x,{sin((0.7*(\x)+3.9)*180/pi)});
\fill  [bottom color=gray!40!white, top color=white,smooth,samples=100,domain=0:10] plot(\x,{sin((0.7*(\x)+3.9)*180/pi)}) -- (10,0) -- (10,24.5) -- (0,24.5) -- (0,0) ;
\fill  [bottom color=white!40!white, top color=white,smooth,samples=100,domain=0:10] plot(\x,{sin((0.7*(\x)+3.9)*180/pi)}) -- (10,0) -- (10,-1.5) -- (0,-1.5) -- (0,0) ;
\draw  (1.1,-1.5)-- (5.6,-1.5);
\draw (5.8,0) node[anchor=north west] {Vapor};
\draw (7.5,27) node[anchor=north west] {Air};
\draw (7.3,23) node[anchor=north west] {Liquid};
\draw [<->]  (10,6)--node[anchor=  south ] {$L_\mathrm{chrst}$} (0,6);
\draw [<->]  (0.5,-1.5)--node[anchor= west]{$H=2.5\times L_\mathrm{chrst}$} (.5,24.5);
\draw [<->]  (0.5,24.5)--node[anchor= west] {$H=0.5\times L_\mathrm{chrst}$} (.5,29);
\draw  [dash pattern=on 2pt off 2pt] (10,-1.5)-- (10,29);
\draw  [dash pattern=on 2pt off 2pt] (0,-1.5)-- (0,29);
\draw[->] (4,0) arc (20:160:0.5) node[anchor=east] {$m^{''}$};
\node[align=center] at (-2, 10) {Periodic};
\node[align=center] at (12, 10) {Periodic};
\node at (5,-2.3 ) { $\alpha=0$, $\vect{U}=(0,0)$, $\cfrac{\mathrm{\partial} p_{}}{\mathrm{\partial} y}=0$, $T=T_{\mathrm{saturat}}+\Delta T_{\mathrm{sup}}$};
\node at (5,30.5 ) {$\cfrac{\mathrm{\partial} \alpha}{\mathrm{\partial} y}=0$,  $\cfrac{\mathrm{\partial} \vect{U}}{\mathrm{\partial} y}=0$, $ p_{\mathrm{}}=P_{\mathrm{saturat}}$, $T=T_{\mathrm{saturat}}$};
\end{scope}
\end{tikzpicture}
\caption{Depiction of the boundary and initial conditions along with the layout of the computational domain employed in the two-dimensional film boiling simulation studies.}
\label{fig:filmboiling}
\end{figure}
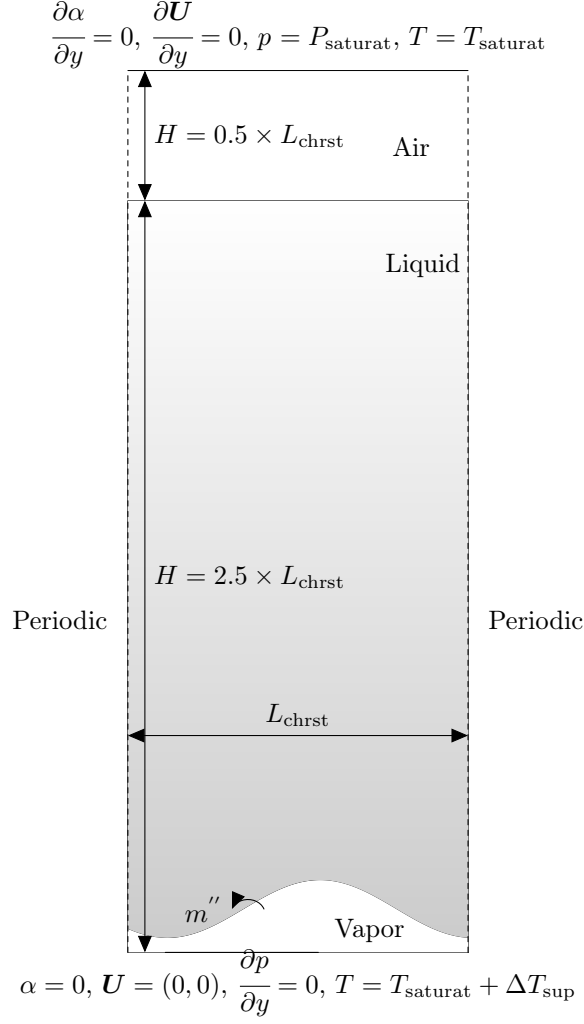

The computational solver's accuracy was verified against the classical Taylor two-dimensional film boiling model for a pure fluid, corroborated by established experimental results. While the prime interest of this study lies in the vertical film boiling phenomena, the adoption of Taylor's model, a standard for analyzing horizontal film boiling, provides a stringent validation protocol. The model was selected due to its widespread acceptance for accuracy, the availability of experimental data for benchmarking purposes, and its applicability to the critical conditions that resemble the vertical film boiling conditions addressed in this work. Key thermophysical properties utilized during the validation phase are itemized in Table~\ref{vertTable}. Simulations were conducted assuming the absence of nanoparticles (\(\varphi=0\)), and the boundary and initial conditions, as well as the geometric specifics of the computational domain, are depicted in Figure~\ref{fig:filmboiling}. Comprehensive descriptions of the benchmark setup and its parameters are documented in the literature \citep{Yahyaee2022ASimulation,Yahyaee2022Level-SetNanofluids,Samkhaniani2017}.

The calculation of the Nusselt number in this 2D film boiling benchmark case adheres to the method established by Esmaeeli and Tryggvason \citep{Esmaeeli2004ComputationsMethod}:

\begin{equation} 
\label{eq:NuNumber}
\text{Nu} = \cfrac{\int_{0}^{L} \left(\cfrac{L_{\text{chrst}}}{\Delta T}\cfrac{\partial T}{\partial y}\Big|_{y=0}\right) dx}{L},
\end{equation}

where \( L_{\text{chrst}} \) represents the characteristic length, defined as follows:

\begin{equation} 
\label{eq:CharacteristicLength}
L_{\text{chrst}} = \sqrt{\cfrac{\sigma}{(\rho_{\text{l}} - \rho_{\text{v}})g}}.
\end{equation}

The derived Nusselt number is then contrasted with the empirical findings of Berenson for two-dimensional film boiling \citep{Berenson1961Film-boilingSurface}:

\begin{equation} 
\label{eq:BerensonNusseltEquation}
\text{Nu} = 0.425\left[\cfrac{\rho_{\text{v}} (\rho_{\text{l}} - \rho_{\text{v}}) g (h_{\text{v}} - h_{\text{l}})}{\lambda_{\text{v}} \mu_{\text{v}} \Delta T} \right].
\end{equation}

\begin{figure}[!htb]
    \centering
    \includegraphics[scale=1]{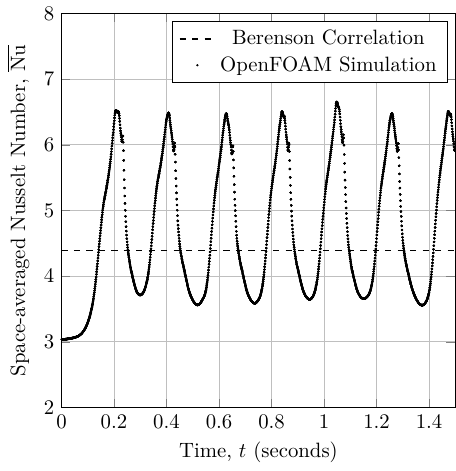}
    \caption{Correlation of Nusselt numbers from numerical simulations with Berenson's empirical formula for 2D film boiling in pure water.}
    \label{fig:NusseltComparison}
\end{figure}

The simulation outcomes, as illustrated in Figure~\ref{fig:NusseltComparison}, yield an average Nusselt number of 4.6 on the surface. This is in close proximity to the 4.25 value derived from Berenson's empirical formula, showcasing a minimal deviation of 7.6\%. Such a comparison endorses the simulation's precision in reflecting the film boiling dynamics as observed in Berenson's empirical experiments.

\begin{figure}[!htb]
\centering
\includegraphics[scale=1]{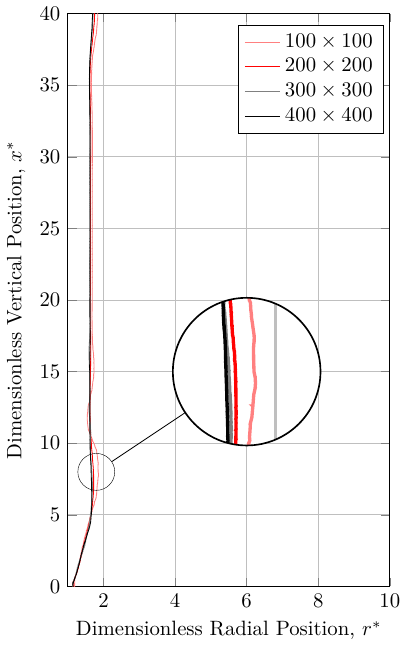}
\caption{Comparison of vapor film interface accuracy at a fixed time across varied grid dimensions \( N_{\text{grids}} = [100 \times 100, 200 \times 200, 300 \times 300, 400 \times 400] \). The \( 300 \times 300 \) grid was chosen for further simulations due to its comparable effectiveness to the \( 400 \times 400 \) grid in accurately capturing the film interface.}
\label{fig:meshindependency}
\end{figure}

To ascertain the most suitable grid size for the computational studies, an extensive mesh sensitivity analysis was performed. This evaluation primarily focused on the precise depiction of the vapor film interface at a designated moment for different mesh sizes. The grid resolutions investigated were \( N_{\text{grids}} = [100 \times 100, 200 \times 200, 300 \times 300, 400 \times 400] \). The findings of this analysis are illustrated in Figure~\ref{fig:meshindependency}. Based on the balance between computational load and precision, the \( 300 \times 300 \) grid was selected for the main simulations, as it offered a performance closely akin to that of the more detailed \( 400 \times 400 \) grid.

\subsection{Dimensionless Representation}
The outcomes of this study are articulated through dimensionless numbers, offering a normalized perspective of the physical phenomena under investigation. This approach facilitates the comparison of results across different scales and conditions.

For radial coordinates, a dimensionless parameter is introduced as:
\begin{equation}
r^* = \cfrac{r}{\delta_\mathrm{rls}},
\end{equation}
where $\delta_\mathrm{rls}$ denotes the reference length scale, itself a derivative from a relevant analytical model \citep{Avramenko2015HeatSurface}:
\begin{equation}
\delta_\mathrm{rls} = \left( \cfrac{4 \lambda_l \mu_l \Delta T x}{g L_{\mathrm{fv}} \rho_l \rho_\mathrm{g}} \right)^{\cfrac{1}{4}}.
\end{equation}

Time is similarly normalized:
\begin{equation}
t^* = \cfrac{\tau}{\mathrm{Pr}}, \quad  \tau = \cfrac{t\mu_\mathrm{g}}{{\delta_\mathrm{rls}}^2 \rho_\mathrm{g}}.
\end{equation}

The temperature field within the domain is normalized with respect to the saturateuration temperature and the wall temperature:
\begin{equation}
T^* = \cfrac{T-T_\mathrm{saturat}}{T_\mathrm{wall}-T_\mathrm{saturat}}.
\end{equation}

The dimensionless Nusselt number, crucial in the assessment of heat transfer, is normalized as per \cite{Avramenko2021UnsteadyBoiling}:
\begin{equation}
\overline{\mathrm{Nu}}= \cfrac{\lambda(\varphi)}{\sqrt{\cfrac{\mu_\mathrm{v,nf}c_\mathrm{p}}{\mathrm{Pr} \lambda_\mathrm{v,nf}{\delta_\mathrm{rls}}^2 }}} \cfrac{\mathrm{\partial} T}{\mathrm{\partial} r}\Big|_{r=R},
\end{equation}
where \( \lambda(\varphi) \) symbolizes the thermal conductivity of the vapor nanofluid phase in proximity to the heating surface.

For comprehensive insight, the spatially-averaged normalized Nusselt number is employed:
\begin{equation}
\overline{\mathrm{Nu}}_{\text{spav}}
= \cfrac{\displaystyle\int_{0}^{\mathrm{L}} \left(\cfrac{\lambda(\varphi)}{\sqrt{\cfrac{\mu_\mathrm{v,nf}c_\mathrm{p}}{\mathrm{Pr} \lambda_\mathrm{v,nf}{\delta_\mathrm{rls}}^2 }}} \cfrac{\mathrm{\partial} T}{\mathrm{\partial} r}\Big|_{r=\mathrm{R}}\right) \,\mathrm{d}r }{\mathrm{L}},
\end{equation}
which integrates the local Nusselt number variations over the cylinder's surface radius \( \mathrm{R} \) and along the length \( \mathrm{L} \) of the vertical plate, thus providing a thorough overview of the heat transfer distribution.

\section{Results and Comparative Analysis}

This study conducts an in-depth examination of the effect of Al\(_2\)O\(_3\) nanoparticle shapes—sphere, brick, blade, cylinder, and platelet—on the dynamics of film boiling along a vertical cylinder, utilizing water-based nanofluids. The primary focus is to assess how these distinct nanoparticle geometries impact the boiling process, specifically in terms of nanoparticle deposition patterns, temperature profiles, and variations in key thermophysical properties such as thermal conductivity and viscosity. One of the pivotal aspects of this analysis is the evaluation of the Nusselt number across different nanoparticle shapes. This metric serves as a crucial indicator of the efficiency of heat transfer enhancements attributable to the nanoparticles.

\subsection{Analysis of Nanoparticle Deposition Dynamics}
\label{subsec:nanoparticleDepositionDynamics}

\begin{figure}[!htb]
\centering
\includegraphics[scale=1]{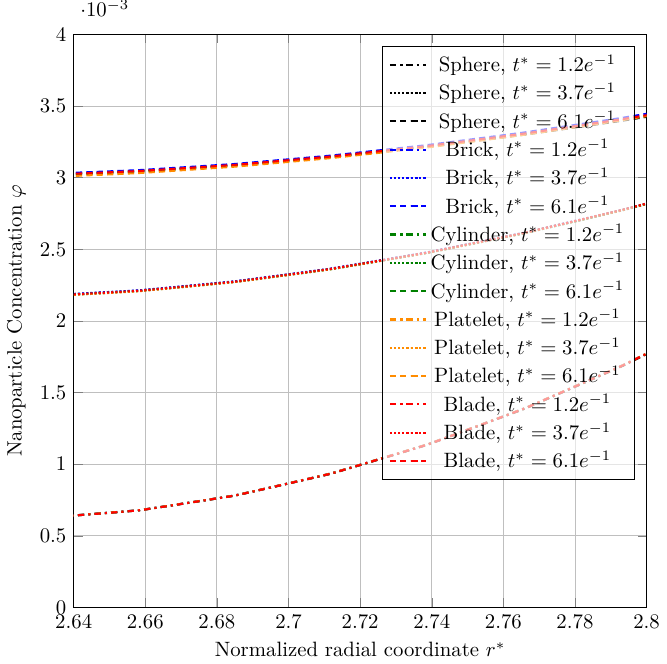}
\caption{Time evolution of the normalized nanoparticle concentration profiles near the heated wall for different nanoparticle shapes, demonstrating the increasing deposition over time during nanofluid boiling. Line probe measurements were taken at a constant axial position ($x^* = 70$) on the cylinder's vertical axis to capture the concentration variations at $t^* = 0.61$.}
\label{fig:nanoparticleConcentrationVsRadius}
\end{figure}

Figure \ref{fig:nanoparticleConcentrationVsRadius} presents the normalized concentration profiles of nanoparticles of various shapes adjacent to the heated wall in a nanofluid boiling scenario. The concentration of nanoparticles in the nanofluid is set to \( 1 \times 10^{-2} \). Interestingly, the data suggests that the differences in the shape of the nanoparticles do not significantly influence their concentration near the wall or their deposition pattern over time. This observation is consistent across three non-dimensional time steps, \( t^* \), indicating that the nanoparticle shape may not be the primary factor influencing the deposition dynamics within the studied regime. 

Notably, as time progresses, there is an apparent increase in the deposition rates, regardless of particle shape. This time-dependent behavior suggests that while the shape of the nanoparticles does not majorly affect the deposition initially, the accumulation of nanoparticles on the surface grows with time. The resulting concentration profiles bear a resemblance to the deposition patterns reported in figure 6 of Malvandi et al. \cite{Malvandi2014ModifiedPipe}, where a low gradient near the wall indicates a near-uniform concentration layer, and the gradient increases progressively as one moves away from the wall.

\subsection{Influence of Nanoparticle Shape on Thermal Conductivity and Viscosity}
\label{subsec:InfluenceNanoparticleShape}

Understanding the influence of nanoparticle shape on the thermal conductivity and viscosity of nanofluids is crucial for the optimization of heat transfer processes. This section discusses the variations in these thermophysical properties, which are determined by the shape of the nanoparticles dispersed within the fluid.

\begin{figure}[!htb]
   \centering
   \includegraphics[scale=1]{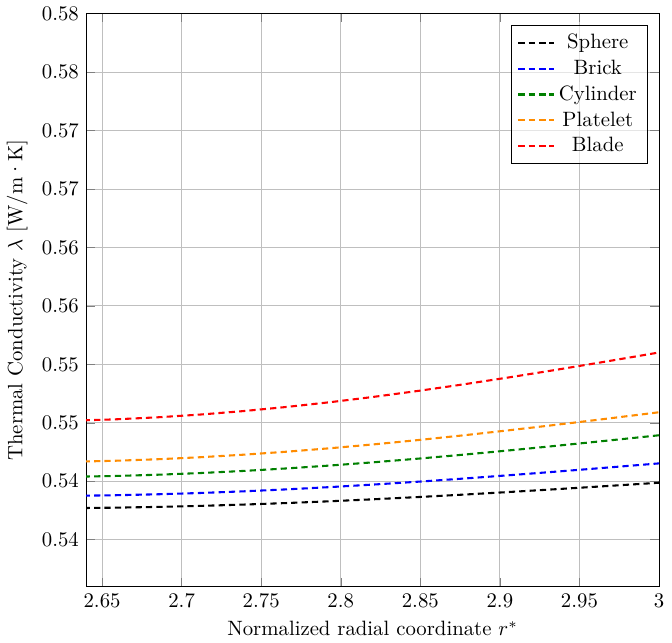}
   \caption{Normalized thermal conductivity profiles across various nanoparticle shapes at a dimensionless time of $t^* = 0.61$. The data was obtained via line probe measurements at a constant axial position ($x^* = 70$) on the cylinder's vertical axis.}
   \label{fig:thermalConductivityComparison}
\end{figure}

The thermal conductivity of nanofluids with differently shaped nanoparticles is captured in Figure~\ref{fig:thermalConductivityComparison}. The calculations are rooted in the Hamilton-Crosser model, where the empirical shape factor, represented as $n$, is inversely proportional to the sphericity, $\Psi$, of the nanoparticles. The sphericity is a dimensionless measure of how closely the shape of the particle approaches that of a perfect sphere. The lower the sphericity, the more pronounced the deviation from spherical shape. This sphericity impacts the effective thermal conductivity of the nanofluid, as depicted in Equation~\ref{eq:revised_thermal_conductivity}, and is detailed for each shape in Table~\ref{table:sphericity_values}.

\begin{figure}[!htb]
   \centering
   \includegraphics[scale=1]{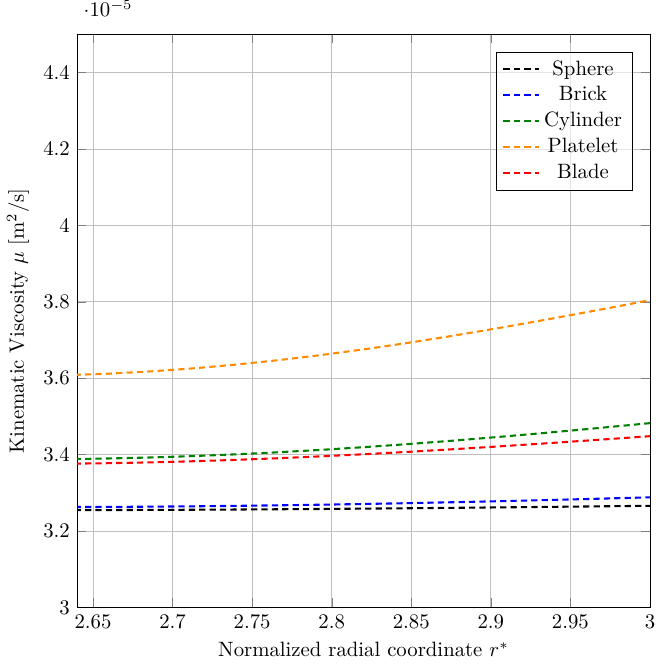}
   \caption{Normalized kinematic viscosity profiles for different nanoparticle shapes at a dimensionless time of $t^* = 0.61$. Measurements were conducted at a fixed axial position ($x^* = 70$) along the cylinder's vertical axis.}
   \label{fig:kinematicViscosityComparison}
\end{figure}

In parallel, the kinematic viscosity of nanofluids, influenced by the shape of nanoparticles, is showcased in Figure~\ref{fig:kinematicViscosityComparison}. Spherical nanoparticles are modeled following Brinkman's equation (Equation~\ref{eq:revised_dynamic_viscosity_spherical}), and for non-spherical nanoparticles, a modified version of the equation (Equation~\ref{eq:revised_dynamic_viscosity_nonspherical}) is utilized, accounting for the shape-induced changes in fluid dynamics.

The analysis indicates that the blade-shaped nanoparticles result in a 
1.38\% increase in thermal conductivity and a 3.73\% rise in kinematic viscosity compared to their spherical counterparts. Similarly, brick-shaped nanoparticles exhibit slight increases of 0.19\% in thermal conductivity and 0.25\% in kinematic viscosity. The cylindrical nanoparticles enhance thermal conductivity by 0.49\% and kinematic viscosity by 4.10\%, while the platelet-shaped nanoparticles show a 0.73\% increase in thermal conductivity and a significant 10.85\% increase in kinematic viscosity.

While the overall effects on thermal conductivity are relatively modest, the substantial variation in kinematic viscosity, especially for platelet-shaped nanoparticles, suggests a noteworthy influence on the fluid's flow behavior and thus its heat transfer capabilities.

\subsection{Temperature Profile Variation Across Nanoparticle Shapes}
\label{subsec:temperatureProfileVariation}

\begin{figure}[!htb]
   \centering
   \includegraphics[scale=1]{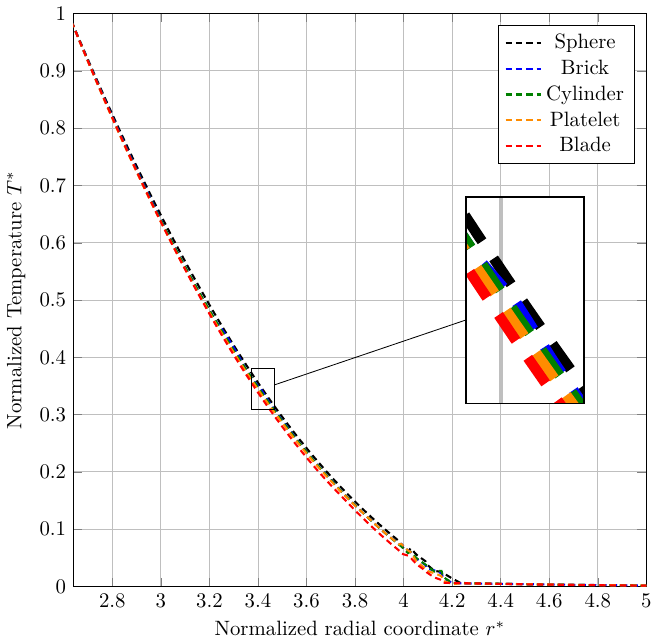}
   \caption{Normalized temperature profiles for nanofluids with different shapes of nanoparticles. Data was obtained via line probe measurements at a constant axial position ($x^* = 70$) on the cylinder's vertical axis at $t^*=0.61$.}
   \label{fig:tempVariation}
\end{figure}

Figure~\ref{fig:tempVariation} illustrates the normalized temperature profiles for different nanoparticle shapes at a constant axial position ($x^* = 70$) on the cylinder's vertical axis during nanofluid boiling. The profiles demonstrate a nearly linear decrease in temperature from the heated wall towards the bulk of the liquid, with a subtle curvature in the middle region resembling the characteristics of temperature profile for nanofluid boiling seen in figure 6 of Malvandi et al. \cite{Malvandi2014ModifiedPipe} or figure 2 of Avramenko et al \cite{Avramenko2021UnsteadyBoiling}.

Among the nanoparticle shapes, the blade-shaped nanoparticles manifest the most pronounced temperature gradient, suggesting a more efficient thermal transport away from the wall. This is followed by platelet and cylinder shapes, which also exhibit significant temperature gradients. In contrast, the sphere and brick shapes present the least steep gradients, indicating lower thermal transport capabilities. The linear-like trend in the temperature profile, with a slight curvature, indicates that while the temperature drops sharply within a short radial distance from the wall, the rate of decrease lessens with increasing distance from the wall. This trend is indicative of a thermal boundary layer development where the thermal conductivity and the shape of nanoparticles play a crucial role.

The interplay between nanoparticle shape and thermal transport is further corroborated by the corresponding thermal conductivity values discussed in Section~\ref{subsec:InfluenceNanoparticleShape}. The steeper gradients associated with blade and platelet shapes align with their higher thermal conductivities, whereas the milder slopes of the sphere and brick shapes correspond to their lower conductivities.

\subsection{Analysis of Space-Averaged Normalized Nusselt Number Variations Across Different Nanoparticle Shapes and Concentrations}
\label{subsec:SpaceAveragedNusseltNumberAnalysis}

This section delves into the influence of nanoparticle shape and concentration (\(\varphi\)) on the space-averaged normalized Nusselt number ($\overline{\mathrm{Nu}}_{\text{spav}}$). At the lower nanoparticle concentration of \(\varphi = 0.01\), the $\overline{\mathrm{Nu}}_{\text{spav}}$ values across different shapes are quite similar, indicating that shape has a less obvious impact on the enhancement of heat transfer.

With increasing nanoparticle concentration to \(\varphi = 0.05\), the $\overline{\mathrm{Nu}}_{\text{spav}}$ profiles start to show a noticeable separation, pointing towards a growing influence of nanoparticle shape on the boiling heat transfer dynamics. This separation is even more distinct at a higher concentration of \(\varphi = 0.1\), confirming a substantial correlation between nanoparticle shape and thermal performance during boiling.

Comparing the space-averaged normalized Nusselt number for different shapes of nanoparticles at \(\varphi = 0.1\) with the pure fluid baseline (\(\varphi = 0\)), the findings are as follows:

\begin{itemize}
    \item Blade-shaped nanoparticles lead to a remarkable increase of approximately 151.77\% in $\overline{\mathrm{Nu}}_{\text{spav}}$, indicating a significant enhancement in heat transfer during boiling.
    \item Brick-shaped nanoparticles register a 84.49\% increase in $\overline{\mathrm{Nu}}_{\text{spav}}$, and cylinder-shaped nanoparticles exhibit a 115.79\% increase.
    \item Platelet-shaped nanoparticles show an increase of about 133.95\%, while sphere-shaped nanoparticles present the least increase of 73.06\% in $\overline{\mathrm{Nu}}_{\text{spav}}$.
\end{itemize}

This observed trend corroborates with the thermal conductivity and temperature distribution data previously discussed (see Figures~\ref{fig:thermalConductivityComparison} and \ref{fig:tempVariation}). Over time, all shapes converge to a steady-state $\overline{\mathrm{Nu}}_{\text{spav}}$.

\begin{figure}[!htb]
   \centering
   \begin{subfigure}[b]{0.45\textwidth}
      \centering
      \includegraphics[scale=0.5]{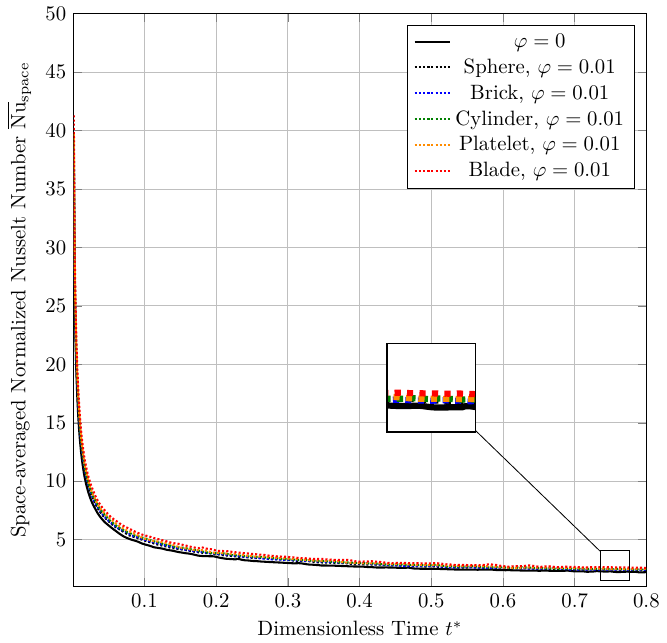}
      \caption{}
      \label{fig:nusseltTimePhi1e-2}
   \end{subfigure}
   \hfill
   \begin{subfigure}[b]{0.45\textwidth}
      \centering
      \includegraphics[scale=0.5]{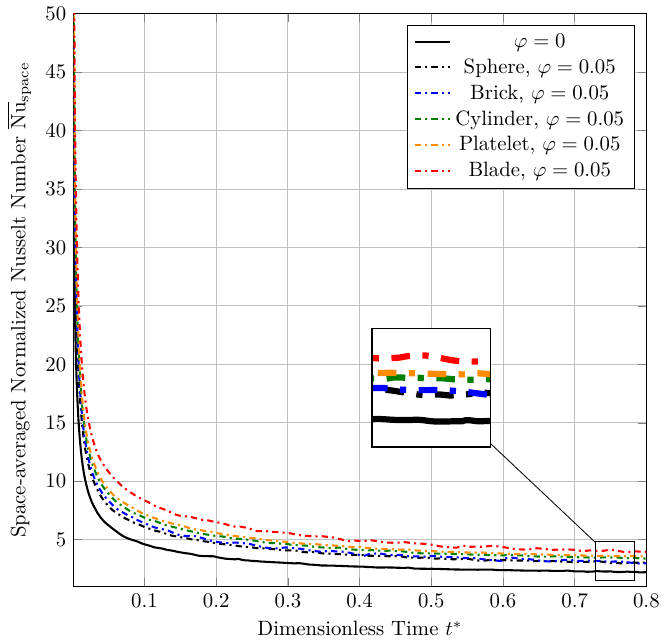}
      \caption{}
      \label{fig:nusseltTimePhi5e-2}
   \end{subfigure}
   \begin{subfigure}[b]{0.45\textwidth}
      \centering
      \includegraphics[scale=0.5]{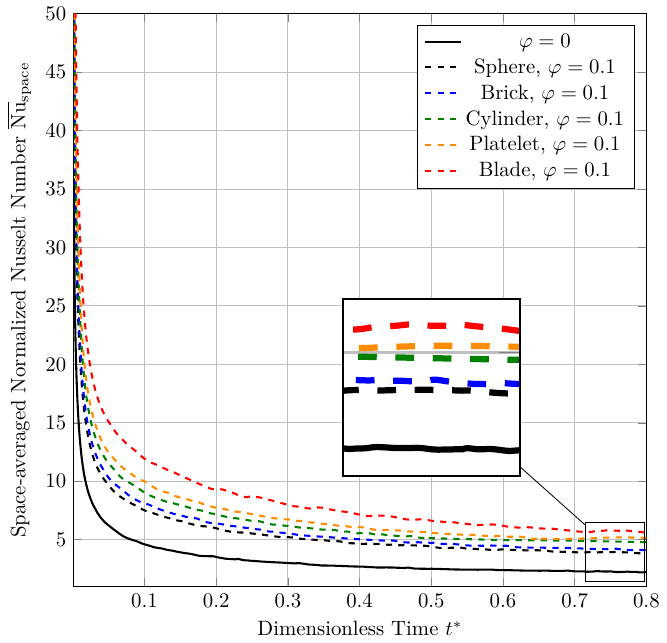}
      \caption{}
      \label{fig:nusseltTimePhi1e-1}
   \end{subfigure} 
   \caption{Space-averaged normalized Nusselt number variation over dimensionless time \(t^*\) for nanofluids with different nanoparticle volume fractions: (a) \(\varphi = 0.01\), (b) \(\varphi = 0.05\), and (c) \(\varphi = 0.1\).}
   \label{fig:nusseltTimeComparison}
\end{figure}

\FloatBarrier
\section{Conclusion}

This investigation methodically evaluates the subtle but critical impact of nanoparticle shapes on the enhancement of nanofluid film boiling along a vertical cylinder. Contrary to initial assumptions, the study reveals that the geometry of alumina (Al$_2$O$_3$) nanoparticles, whether spherical, brick, blade, cylinder, or platelet, has only a minimal effect on deposition dynamics at the nanoscale level. Nonetheless, an observed increase in nanoparticle deposition rate over time suggests that prolonged interaction leads to greater surface accumulation, supporting findings from prior research.

It is particularly noteworthy that while the effect of nanoparticle shape on thermal conductivity is moderate, the shape significantly alters the nanofluid's kinematic viscosity. This is most evident with platelet-shaped nanoparticles, which markedly affect fluid movement and thus the boiling heat transfer mechanism. Among the shapes studied, blade-shaped nanoparticles stand out, significantly improving thermal transport as indicated by the sharp temperature gradients and increased thermal conductivity, consistent with theoretical predictions.

Furthermore, the research highlights the space-averaged normalized Nusselt number as a vital indicator, with increased nanoparticle concentration accentuating the differences between shapes. Blade-shaped nanoparticles, for example, demonstrate a remarkable enhancement in boiling heat transfer efficiency, surpassing other shapes by a substantial margin.

This work lays the groundwork for optimizing the application of nanofluids in film boiling, particularly in industries where thermal management and safety are paramount. Employing the Continuous-Species-Transfer (CST) methodology within a Computational Multi-Fluid Dynamics (CMFD) framework has proven instrumental in simulating the intricate behaviors of nanofluid flow and heat transfer, providing key insights for future empirical and computational exploration.

In summary, the research elucidates that nanoparticle shape profoundly influences the thermal transport characteristics. These insights are crucial for advancing heat transfer processes in scenarios requiring high thermal fluxes. As technological advancements push the boundaries of nanofluid applications, a detailed comprehension of these factors becomes increasingly important for the design and management of high-performance thermal systems.

\clearpage

\bibliography{Bibliography}

\clearpage

\appendix
\renewcommand{\thesection}{\Alph{section}} 
\setcounter{section}{0} 

\renewcommand{\theequation}{\thesection.\arabic{equation}} 
\setcounter{equation}{0} 
\section*{Appendix}
\section{Nanoparticle Dynamics Equation Derivation} \label{apendixNanoCon}

\addcontentsline{toc}{section}{Appendix}
\label{subsec:DerivationNanoparticleConcentration}

The equation governing the behavior of nanoparticles in terms of their volumetric concentration across vapor, liquid nanofluid states, and at their interface is formulated by extending the nanoparticle concentration equation commonly applied within a single phase. The comprehensive equation (Equation~\ref{eq:final_governing_equationAA}), derived from existing theoretical models \citep{Michaelides2015BrownianLiquids,Malvandi2016ThermophoresisCylinder,Lin2018EffectsSimulation}, is given by:

\begin{equation}
    \cfrac{\partial \phi}{\partial t} + \nabla \cdot (\phi \mathbf{U}) = \nabla \cdot (D_\mathrm{Bn} \nabla \phi) + \nabla \cdot\left[D_\mathrm{Thps} \cfrac{\nabla T}{T_{\infty}}\right],
\end{equation}
where the terms \(D_\mathrm{Bn}\) and \(D_\mathrm{Thps}\) represent the diffusion coefficients for Brownian motion and thermophoresis, respectively. To describe these phenomena, the coefficients are mathematically defined as:

\begin{equation}
    D_\mathrm{Bn} = \cfrac{k_\mathrm{Bltzmn} T}{3 \pi \mu_{\text{bf}} d_p},
\label{eq:brownianD}
\end{equation}

\begin{equation}
    D_\mathrm{Thps} = \beta \cfrac{\mu_{\text{bf}}}{\rho_{\text{bf}}} \phi,
\label{eq:thermophoresisD}
\end{equation}
with \(\beta\) being a proportional constant expressed as \(0.26 \lambda_{\text{bf}} /(2 \lambda_{\text{bf}} + \lambda_{\text{np}})\). To simplify the modeling, these coefficients are further reduced to temperature-dependent forms:

\begin{equation}
    D_\mathrm{Bn} = D_\mathrm{Bn}' T, \quad D_\mathrm{Thps} = D_\mathrm{Thps}' T_{\infty} \phi.
\end{equation}

For the liquid nanofluid phase, the governing equation is articulated as:

\begin{align}
    \cfrac{\partial \alpha_{\mathrm{l,nf}} \overline{\phi}^{\mathrm{l,nf}}}{\partial t} 
    &+ \nabla \cdot (\alpha_{\mathrm{l,nf}} \overline{\phi}^{\mathrm{l,nf}} \overline{\mathbf{U}}^{\mathrm{l,nf}}) \nonumber \\
    &= \nabla \cdot (\alpha_{\mathrm{l,nf}} \overline{D_\mathrm{Bn}'}^{\mathrm{l,nf}} \overline{T}^{\mathrm{l,nf}} \nabla \overline{\phi}^{\mathrm{l,nf}}) \nonumber \\
    &+ \nabla \cdot(\alpha_{\mathrm{l,nf}} \overline{D_\mathrm{Thps}'}^{\mathrm{l,nf}} \overline{\phi}^{\mathrm{l,nf}} \nabla T^{\mathrm{l,nf}}),
\label{eq:conditioning_volume_averaging2}
\end{align}

Equation~\ref{eq:conditioning_volume_averaging2} reveals the dynamics of nanoparticle distribution within the liquid phase, accounting for both the random motion due to Brownian diffusion and the directed movement as a result of thermophoretic effects.

Similarly, for the vapor nanofluid phase, the equation becomes:

\begin{align}
    \cfrac{\partial \alpha_{\mathrm{v,nf}} \overline{\phi}^{\mathrm{v,nf}}}{\partial t} 
    &+ \nabla \cdot (\alpha_{\mathrm{v,nf}} \overline{\phi}^{\mathrm{v,nf}} \overline{\mathbf{U}}^{\mathrm{v,nf}}) \nonumber \\
    &= \nabla \cdot (\alpha_{\mathrm{v,nf}} \overline{D_\mathrm{Bn}'}^{\mathrm{v,nf}} \overline{T}^{\mathrm{v,nf}} \nabla \overline{\phi}^{\mathrm{v,nf}}) \nonumber \\
    &+ \nabla \cdot(\alpha_{\mathrm{v,nf}} \overline{D_\mathrm{Thps}'}^{\mathrm{v,nf}} \overline{\phi}^{\mathrm{v,nf}} \nabla T^{\mathrm{v,nf}}).
\label{eq:conditioning_volume_averagingVapor}
\end{align}

Equation~\ref{eq:conditioning_volume_averagingVapor} delineates the behavior of nanoparticles in the vapor phase, a critical consideration for comprehensive modeling of the nanofluid dynamics during boiling.

By amalgamating Equations~\ref{eq:conditioning_volume_averaging2} and \ref{eq:conditioning_volume_averagingVapor}, a more robust model emerges. In scenarios where velocity by gravitational forces overshadow the velocity by phase change, certain surface integral terms within the governing equations can be disregarded. This assumption, which is supported by the literature \citep{Avramenko2015HeatSurface,Malvandi2016ThermophoresisCylinder}, simplifies the model without compromising its integrity for the specific case of film boiling influenced predominantly by gravity. Thus, the model remains faithful to the physics of the problem while achieving computational efficiency.

In the quest for a more holistic understanding of nanofluid dynamics, this study introduces \(\varphi\), a parameter encapsulating the overall concentration of nanoparticles across both liquid and vapor phases in the computational domain. Defined as a combination of contributions from each phase, \(\varphi\) serves as a comprehensive measure of nanoparticle concentration:

\begin{equation}
    \varphi = \alpha_{\mathrm{l,nf}} \overline{\phi}^{\mathrm{l,nf}} + (1-\alpha_{\mathrm{l,nf}}) \overline{\phi}^{\mathrm{v,nf}}.
\label{eq:totalConcentration}
\end{equation}

This work further delves into refining the term addressing Brownian dynamics in Equation~\ref{eq:conditioning_volume_averaging2}. The objective is to reframe it in terms of \(\varphi\) rather than \(\phi\), leading to a more integrated understanding of nanoparticle behavior. The revised formulation breaks down the term into three components, each representing a different aspect of the nanofluid's behavior:

\begin{equation}
    \begin{aligned}
        & \nabla \cdot\left( \sum_{k=\mathrm{nf,v}} \nabla \cdot (\alpha_{\mathrm{k}} \overline{D_\mathrm{Bn}'}^{\mathrm{k}} \overline{T}^{\mathrm{k}} \nabla \overline{\phi}^{\mathrm{k}}) \right) = \underbrace{\nabla \cdot\left( \sum_{k=\mathrm{nf,v}} \nabla\left( \overline{D_\mathrm{Bn}'}^{\mathrm{k}} \overline{T}^{\mathrm{k}} \alpha_k \overline{\phi}^k \right) \right)}_{\text{(a)}} \\
        & - \underbrace{\nabla \cdot\left( \sum_{k=\mathrm{nf,v}} \overline{D_\mathrm{Bn}'}^{\mathrm{k}} \overline{T}^{\mathrm{k}} \overline{\phi}^k \nabla \alpha_k \right)}_{\text{(b)}} \\
        & - \underbrace{\nabla \cdot\left( \sum_{k=\mathrm{nf,v}} \overline{D_\mathrm{Bn}'}^{\mathrm{k}} \alpha_k \overline{\phi}^k \nabla \overline{T}^{\mathrm{k}} \right)}_{\text{(c)}} .
    \end{aligned}
\label{eq:bulknterfacial_contributions}
\end{equation}

To address component (a) in Equation~\ref{eq:bulknterfacial_contributions}, a methodological approximation inspired by Marschall et al. \cite{Marschall2012NumericalOpenFOAM} is employed. This involves integrating the molar mixture concentration \(S_i\) into the equation to better represent the blended effects of nanoparticle concentration across the interface:

\begin{equation}
S_i = \cfrac{{\overline{D_i}}^{\omega} \alpha_{\omega} {\overline{s_i}}^{\omega} + {\overline{D_i}}^\epsilon \alpha_{\epsilon} {\overline{s_i}}^\epsilon}{\alpha_{\omega} {\overline{D_i}}^{\omega} + \alpha_\epsilon {\overline{D_i}}^\epsilon}.
\label{eq:molarMixtureConcentration}
\end{equation}

For an effective approximation of \(\varphi\), the following equation is utilized:

\begin{equation}
    \varphi \approx \cfrac{{\overline{D_\mathrm{Bn}}}^{\text{l,nf}} \alpha_{\text{l,nf}} {\overline{\phi}}^{\text{l,nf}} + {\overline{D_\mathrm{Bn}}}^{\text{v,nf}} \alpha_{\text{v,nf}} {\overline{\phi}}^{\text{v,nf}}}{\alpha_{\text{l,nf}} {\overline{D_\mathrm{Bn}}}^{\text{l,nf}} + \alpha_{\text{v,nf}} {\overline{D_\mathrm{Bn}}}^{\text{v,nf}}}.
\label{eq:Ci_approximation2}
\end{equation}

Within cells fully occupied by either nanofluid or vapor, the average temperatures \( \overline{T}^{\mathrm{l,nf}} \) and \( \overline{T}^{\mathrm{v,nf}} \) can be represented by a singular temperature value \( T \), as posited in literature \citep{Habla2016DevelopmentStage}. In film boiling scenarios, equating \( \overline{T}^{\mathrm{l,nf}} \) to \( \overline{T}^{\mathrm{v,nf}} \) simplifies the equation, leading to a more manageable form:

\begin{equation}
 \alpha_{\text{l,nf}} {\overline{D_\mathrm{Bn}'}}^{\text{l,nf}} \overline{T}^{\text{l,nf}} + \alpha_{\text{v,nf}} {\overline{D_\mathrm{Bn}'}}^{\text{v,nf}} \overline{T}^{\text{v,nf}}=D_\mathrm{Bn}' T.
\end{equation}

The above simplification enables the restructuring of Equation \ref{eq:Ci_approximation2} and the subsequent adaptation of term (a) in Equation \ref{eq:bulknterfacial_contributions} to reflect \( \varphi \). The resulting expression is as follows:

\begin{align}
    \sum_{k=\text{l,nf and v,nf}} \nabla \cdot \left( \nabla \left( {\overline{D_\mathrm{Bn}' }}^k \overline{T}^k \alpha_k {\overline{\phi}}^k \right) \right) &= \nabla \cdot \left( \nabla \left( D_\mathrm{Bn}' T \varphi \right) \right) \nonumber \\
    &+ \nabla \cdot \left( D_\mathrm{Bn}'T \nabla \varphi \right) \nonumber \\
    &+ \nabla \cdot \left( \varphi T \nabla D_\mathrm{Bn}'  \right) \nonumber \\
    &+ \nabla \cdot \left( \varphi D_\mathrm{Bn}' \nabla T  \right).
\label{eq:revisedBulkInterfacialContributions}
\end{align}

The exploration of term (b) in Equation \ref{eq:bulknterfacial_contributions} centers on the interfacial interactions between the nanofluid and vapor phases. By recognizing the inverse relationship between the liquid nanofluid (\(\mathrm{l,nf}\)) and vapor (\(\mathrm{v,nf}\)) phase volume fractions, namely \(\alpha_{\mathrm{l,nf}}=1-\alpha_{\mathrm{v,nf}}\) and \(\nabla \alpha_{\mathrm{l,nf}}=-\nabla \alpha_{\mathrm{v,nf}}\) at the interface, term (b) is articulated as:

\begin{equation}
\nabla \cdot \left[ \left( \overline{D_\mathrm{Bn}' }^{\mathrm{l,nf}} \overline{T}^{\mathrm{l,nf}}\overline{\phi}^{\mathrm{l,nf}} - \overline{D_\mathrm{Bn}' }^{\mathrm{v,nf}} \overline{T}^{\mathrm{v,nf}}\overline{\phi}^{\mathrm{v,nf}} \right) \nabla \alpha_{\mathrm{l,nf}} \right].
\end{equation}

For the film boiling scenario, it is feasible to employ the simplification \( \overline{T}^{\mathrm{l,nf}} = \overline{T}^{\mathrm{v,nf}} \), thus considering a unified temperature \( T \) for both phases. An important aspect of this analysis is the application of Henry's law, which quantifies the differential solubility of nanoparticles in the liquid and vapor phases through the distribution coefficient \( \mathrm{H}_\mathrm{cnst} \). This coefficient, often larger than 1, signifies a higher nanoparticle concentration in the liquid phase compared to the vapor phase. For this study, \( \mathrm{H}_\mathrm{cnst} \) is assumed to be 8.

The introduction of Henry's law is pivotal in defining the concentration jump condition at the vapor-liquid interface. This condition, depicted in Equation~(\ref{eq:jump_conditions}), captures the concentration discrepancy across the interface:

\begin{equation}
    \mathrm{H}_\mathrm{cnst} = \cfrac{\phi_{I,\mathrm{l,nf}}}{\phi_{I,\mathrm{v,nf}}},
    \label{eq:jump_conditions}
\end{equation}

Incorporating Henry's law into the interfacial dynamics, as represented by Equation~(\ref{eq:jump_conditions}), allows for a refined expression of term (b) in Equation \ref{eq:bulknterfacial_contributions}. This results in a formulation that takes into account the relative nanoparticle solubility in both phases:

\begin{equation}
\nabla \cdot \left[ \left( \overline{D_\mathrm{Bn}' }^{\mathrm{l,nf}} - \cfrac{\overline{D_\mathrm{Bn}' }^{\mathrm{v,nf}}}{\mathrm{H}_\mathrm{cnst}} \right) T \overline{\phi}^{\mathrm{l,nf}} \nabla \alpha_{\mathrm{l,nf}} \right].
\label{eq:Ci_relation}
\end{equation}

Finally, the incorporation of Henry's law, coupled with Equation \ref{eq:Ci_relation}, leads to a comprehensive expression for the overall concentration \( \varphi \), effectively capturing the nanoparticle transport dynamics across the fluid interface:

\begin{equation}
\nabla \cdot \left[ \left( \cfrac{\overline{D_\mathrm{Bn}' }^{\mathrm{l,nf}} - \overline{D_\mathrm{Bn}' }^{\mathrm{v,nf}} / \mathrm{H}_\mathrm{cnst}}{\alpha_{\mathrm{l,nf}} + (1 - \alpha_{\mathrm{l,nf}}) / \mathrm{H}_\mathrm{cnst}} \right) T \varphi \nabla \alpha_{\mathrm{l,nf}} \right].
\end{equation}

The investigation of term (c) in Equation \ref{eq:bulknterfacial_contributions} sheds light on how thermal gradients impact nanoparticle dynamics, originally expressed as:

\begin{equation}
\nabla \cdot \left[ \left( \alpha_{\mathrm{l,nf}}\overline{D_\mathrm{Bn}' }^{\mathrm{l,nf}} \overline{\phi}^{\mathrm{l,nf}} \nabla \overline T ^{\mathrm{l,nf}}+ \alpha_{\mathrm{v,nf}}\overline{D_\mathrm{Bn}' }^{\mathrm{v,nf}} \overline{\phi}^{\mathrm{v,nf}}\nabla \overline T ^{\mathrm{v,nf}} \right)  \right].
\end{equation}

In regions solely populated by either the nanofluid or vapor, the temperature gradients of the respective phases can be approximated by \( \nabla T \). In the context of film boiling on a vertical surface, this approximation assumes parallelism in direction and magnitude between \( \nabla \overline{T}^{\mathrm{l,nf}} \), \( \nabla \overline{T}^{\mathrm{v,nf}} \), and \( \nabla T \). Leveraging Henry's law at the interface, term (c) simplifies to:

\begin{equation}
\nabla \cdot \left[ \left( \alpha_{\mathrm{l,nf}}\overline{D_\mathrm{Bn}' }^{\mathrm{l,nf}} + \cfrac{\alpha_{\mathrm{v,nf}}\overline{D_\mathrm{Bn}' }^{\mathrm{v,nf}}}{\mathrm{H}_\mathrm{cnst}} \right) \overline{\phi}^{\mathrm{l,nf}} \nabla T \right].
\end{equation}

Subsequently, incorporating Henry's law enables a reformulation of this term in relation to the overall concentration \( \varphi \):

\begin{equation}
\nabla \cdot \left[ \left( \cfrac{\alpha_{\mathrm{l,nf}}\overline{D_\mathrm{Bn}' }^{\mathrm{l,nf}} +(1 - \alpha_{\mathrm{l,nf}}) \overline{D_\mathrm{Bn}' }^{\mathrm{v,nf}} / \mathrm{H}_\mathrm{cnst}}{\alpha_{\mathrm{l,nf}} + (1 - \alpha_{\mathrm{l,nf}}) / \mathrm{H}_\mathrm{cnst}} \right) \varphi \nabla T \right].
\end{equation}

The term corresponding to thermophoresis in the governing equation (Equation \ref{eq:conditioning_volume_averaging2}) is initially framed as:

\begin{equation}
\nabla \cdot \left[ \left( \alpha_{\mathrm{l,nf}}\overline{D_\mathrm{Thps}' }^{\mathrm{l,nf}} \overline{\phi}^{\mathrm{l,nf}} + \alpha_{\mathrm{v,nf}}\overline{D_\mathrm{Thps}' }^{\mathrm{v,nf}} \overline{\phi}^{\mathrm{v,nf}} \right) \nabla T \right].
\end{equation}

Application of Henry's law at the fluid boundary simplifies this expression:

\begin{equation}
\nabla \cdot \left[ \left( \alpha_{\mathrm{l,nf}}\overline{D_\mathrm{Thps}' }^{\mathrm{l,nf}} + \cfrac{\alpha_{\mathrm{v,nf}}\overline{D_\mathrm{Thps}' }^{\mathrm{v,nf}}}{\mathrm{H}_\mathrm{cnst}} \right) \overline{\phi}^{\mathrm{l,nf}} \nabla T \right].
\end{equation}

Further extending Henry's law's application, the thermophoresis term is expressed in terms of \( \varphi \):

\begin{equation}
\nabla \cdot \left[ \left( \cfrac{\alpha_{\mathrm{l,nf}}\overline{D_\mathrm{Thps}' }^{\mathrm{l,nf}} + (1 - \alpha_{\mathrm{l,nf}}) \overline{D_\mathrm{Thps}' }^{\mathrm{v,nf}} / \mathrm{H}_\mathrm{cnst}}{\alpha_{\mathrm{l,nf}} + (1 - \alpha_{\mathrm{l,nf}}) / \mathrm{H}_\mathrm{cnst}} \right) \varphi \nabla T \right].
\end{equation}

The synthesis of our detailed analytical process culminates in a unified governing equation. This equation seamlessly integrates the diverse transport mechanisms of nanoparticles and aptly characterizes the dynamics at the fluid interface in film boiling scenarios. The definitive form of the governing equation is:

\begin{align}
    \cfrac{\partial \varphi}{\partial t} + \nabla \cdot \left( \varphi \mathbf{U} \right) &= \nabla \cdot \left( D_\mathrm{Bn}'T \nabla \varphi \right) + \nabla \cdot \left( \varphi T \nabla D_\mathrm{Bn}'  \right) \nonumber \\
    &\quad + \nabla \cdot \left( \varphi D_\mathrm{Bn}' \nabla T  \right) \nonumber \\
    &\quad - \nabla \cdot \left[ \left( \cfrac{\overline{D_\mathrm{Bn}' }^{\mathrm{l,nf}} - \overline{D_\mathrm{Bn}' }^{\mathrm{v,nf}} / \mathrm{H}_\mathrm{cnst}}{\alpha_{\mathrm{l,nf}} + (1 - \alpha_{\mathrm{l,nf}}) / \mathrm{H}_\mathrm{cnst}} \right) \varphi T \nabla \alpha_{\mathrm{l,nf}} \right] \nonumber \\
    &\quad - \nabla \cdot \left[ \left( \cfrac{\alpha_{\mathrm{l,nf}}\overline{D_\mathrm{Bn}' }^{\mathrm{l,nf}} +(1 - \alpha_{\mathrm{l,nf}}) \overline{D_\mathrm{Bn}' }^{\mathrm{v,nf}} / \mathrm{H}_\mathrm{cnst}}{\alpha_{\mathrm{l,nf}} + (1 - \alpha_{\mathrm{l,nf}}) / \mathrm{H}_\mathrm{cnst}} \right) \varphi \nabla T \right]\nonumber \\
    &\quad + \nabla \cdot \left[ \left( \cfrac{\alpha_{\mathrm{l,nf}}\overline{D_\mathrm{Thps}' }^{\mathrm{l,nf}} +(1 - \alpha_{\mathrm{l,nf}}) \overline{D_\mathrm{Thps}' }^{\mathrm{v,nf}} / \mathrm{H}_\mathrm{cnst}}{\alpha_{\mathrm{l,nf}} + (1 - \alpha_{\mathrm{l,nf}}) / \mathrm{H}_\mathrm{cnst}} \right) \varphi \nabla T \right].
    \label{eq:final_governing_equation}
\end{align}

Equation \ref{eq:final_governing_equation} embodies a comprehensive model that is applicable across the entire computational domain. It adeptly represents the behavior of nanoparticles in both the nanofluid and vapor phases, as well as their interaction at the interface. This governing equation adeptly captures the standard nanoparticle transport phenomena, along with the specialized interface conditions, and addresses the contrasting properties of the two phases in the context of film boiling.

\end{document}